\documentclass{article}
\usepackage{graphicx}
\usepackage{amsthm}
\usepackage{enumitem}
\graphicspath{{figures/}}

\theoremstyle{plain}
\newtheorem{example}{Example}
\newtheorem{remark}{Remark}
\newtheorem{assumption}{Assumption}
\newtheorem{method}{Method}
\newtheorem{theorem}{Theorem}
\newtheorem{definition}{Definition}
\usepackage{amsmath,amssymb,amsfonts}
\usepackage{xcolor}
\usepackage[linesnumbered]{algorithm2e}
\usepackage{appendix}
\newcommand{\email}[1]{\texttt{#1}}

\newcommand{\R}{\mathbb{R}}
\newcommand{\N}{\mathbb{N}}
\newcommand{\xmin}{x_\mathrm{min}}
\newcommand{\xmax}{x_\mathrm{max}}
\newcommand{\xinf}{x_\mathrm{inf}}
\newcommand{\xsup}{x_\mathrm{sup}}

\newcommand{\bfX}{\mathbf{X}}
\newcommand{\bfw}{\mathbf{w}}
\newcommand{\bflambdaw}{\boldsymbol{\lambda}_{\mathrm{w}}}
\newcommand{\bflambdawopt}{\boldsymbol{\lambda}_{\mathrm{w, opt}}}
\newcommand{\lambdar}{\lambda_{\mathrm{r}}}
\newcommand{\lambdaropt}{\lambda_{\mathrm{r, opt}}}
\newcommand{\bfmu}{\boldsymbol{\mu}}
\newcommand{\bfSigma}{\boldsymbol{\Sigma}}
\newcommand{\bfnabla}{\boldsymbol{\nabla}}

\newcommand{\bfsigma}{\boldsymbol{\sigma}}
\newcommand{\bfrho}{\boldsymbol{\rho}}
\usepackage{url}
\newcommand{\one}{\mathbf{1}}

\newcommand{\ouralgorithm}{ $\Lambda$-Newton-Bis}

\newcommand{\Review}[1]{{#1}}
\usepackage{booktabs}

\usepackage{hyperref}
\begin{document}

\date{}
\title{Numerical methods for lambda quantiles: \\robust evaluation and portfolio optimisation\thanks{Accepted for publication in SIAM Journal on Financial Mathematics}}

\author{
I. Peri\thanks{ Birkbeck Business School, Birkbeck University of London, Malet St, Bloomsbury, London WC1E 7HX, United Kingdom; \email{I.Peri@bbk.ac.uk}}
\and 
Linus Wunderlich\thanks{School of Mathematical Sciences, Queen Mary University of London, Mile End Road, London E1 4NS, United Kingdom, \email{L.Wunderlich@qmul.ac.uk}}
}
\maketitle
\begin{abstract}
Lambda quantiles, originally introduced as lambda value at risk, generalise the classical value at risk by allowing for a variable confidence level.  
This work presents efficient algorithms for computing lambda quantiles and demonstrates their application in portfolio optimisation. We first develop a robust algorithm, $\Lambda$-Newton-Bis, that combines Newton’s method with a bisection strategy to ensure global convergence. The algorithm handles potential discontinuities and achieves local quadratic convergence under standard regularity assumptions. To address cases with multiple roots, we also propose an interval analysis approach. We then demonstrate the algorithm’s computational efficiency and practical relevance within a portfolio optimization framework. To this end, we develop two alternative solution methods that incorporate the $\Lambda$-Newton-Bis procedure. Numerical experiments confirm the algorithm's convergence properties and highlight its computational advantages in optimization tasks based on lambda quantiles.
\end{abstract}


\section{Introduction}

In finance and portfolio management, having a fast and reliable method for computing risk measures plays an important role, especially when analytical solutions are unavailable or when optimizing risk measures is a central goal, such as in portfolio allocation. Traditional risk measures, while widely adopted, often fail to capture nuanced risk profiles, especially for assets with heavy-tailed distributions. One innovative alternative that may address these limitations is the lambda quantile. However, the practical use of lambda quantiles necessitates efficient numerical solutions, an area that remains underexplored in the existing literature.

Lambda quantiles~\cite{fmp14}, originally introduced as lambda value at risk ($\Lambda$VaR)   within the risk management context, were later termed \emph{lambda quantile} or \emph{$\Lambda$-quantile}~\cite{bp19} to highlight their broader application as a generalized quantile. Unlike traditional quantiles, lambda quantiles use a variable confidence level function $\Lambda$, dependent on the profits and losses of an asset. This may allow to better capture tail risks, discriminate among assets with identical quantile, i.e. value at risk (VaR), and associate more risk to assets with heavy-tailed distributions of profits and losses. These features make lambda quantiles particularly appealing in volatile or uncertain market conditions.

The literature on lambda quantiles focuses on the theoretical properties that qualify them as suitable risk measures. The foundational study~\cite{fmp14} highlighted their monotonicity and quasiconvexity on distributions. Further research~\cite{bpr17} explored additional properties relevant for applications and backtesting, such as robustness of the historical estimator, elicitability, and consistency. In \cite{bp19} locality was identified  as a key property for their axiomatisation, formalizing the idea that, as with traditional quantiles, shifting probability mass to the left or right of the lambda quantile does not alter its value. Although lambda \Review{quantiles} lack cash-additivity and quasiconvexity when applied to random variables, subsequent work~\cite{han2021cash} demonstrated their cash-subadditivity and quasi-star-shapedness, aligning well with the needs for capital requirement accounting for ambiguous interest rates and liquidity risk. Furthermore, the recent study~\cite{chambers2024max} demonstrated lambda quantiles' max-stability under first stochastic dominance.

Recent studies have enhanced the applicability of lambda quantiles in finance and insurance, addressing the issue of risk contributions~\cite{ince:22}, the risk-sharing problem under both homogeneous beliefs~\cite{liu2024risk,xia2024optimal} and heterogeneous beliefs~\cite{liu2024brisk}. Additionally, the optimal insurance problem based on lambda quantiles has been recently investigated~\cite{boonen2024optimal}, building on the earlier work~\cite{balbas2023lambda}. Robust versions of lambda quantiles, as the worst-case and best-case under uncertainty, have been studied recently and applied in portfolio selection~\cite{han2024robust}. 

The concept of quantiles with a variable confidence level has inspired further studies. For instance, \cite{bignozzi2020risk} developed a cash additive risk measure based on benchmark loss distributions, while \cite{balbas2023lambda} proposed the lambda-fixed point risk measure as an extension of lambda quantiles. Despite this expanding literature highlighting the robust theoretical foundations and versatility of lambda quantiles, addressing their computational challenges remains an open problem. 

While theoretical investigations of lambda quantiles abound, efforts to develop robust and efficient computational methods are limited. Early empirical studies, such as~\Review{\cite{hmp18}} demonstrated their potential for assessing equity risk under various market conditions by regularly recalibrating the $\Lambda$ function. This approach was refined~\cite{corbetta2018backtesting}, who introduced a backtesting framework for lambda quantiles. Later,~\cite{ince:22} suggested a methodology to calculate lambda quantiles and their risk contributions using kernel density estimation. Despite these advancements, a definitive computational framework for lambda quantiles is still lacking. Additionally, previous studies have relied on general-purpose equation solvers such as Matlab's \texttt{fsolve} or Python's SciPy package, which often lack robustness and frequently encounter convergence issues.

In this paper, we address this gap by introducing an ad-hoc numerical method to compute lambda quantiles. Our proposed method combines the Newton method, known for its fast local quadratic convergence, with the bisection method, which addresses the major failure issues of the Newton method, such as oscillation and divergence. This hybrid\ouralgorithm{} algorithm ensures global convergence under only mild conditions and achieves faster local quadratic convergence with minimal additional assumptions. Practical examples illustrate the algorithm's robustness across a variety of distribution functions, including discontinuous cases.

Finally, we explore the application of the\ouralgorithm{} algorithm to portfolio optimization. Specifically, we frame a portfolio allocation problem where the objective is to minimize the lambda value at risk under a return constraint. Two solution methods are proposed: one based on the Karush-Kuhn-Tucker (KKT) conditions and the other on a penalty method. We compare the performance of these two methods and the efficiency of the $\Lambda$-Newton-Bis algorithm in this particular setting by using numerical experiments on two-asset and three-asset portfolios.

The paper is structured as follows: Section \ref{sec:LQ} reviews some relevant properties of lambda quantiles; Section \ref{sec:NewtonBis} presents the $\Lambda$-Newton-Bis algorithm and states the convergence results; Section \ref{sec:PortLQ} discusses the portfolio optimisation framework; Section \ref{sec:conclusion} concludes; Appendix \ref{sec:appendix_proofs} contains proofs of the theoretical results.

\section{Lambda Quantiles}\label{sec:LQ}

Consider an atomless probability space $(\Omega, \mathcal{F}, \mathbb{P})$. Let $\mathcal{X}$ denote the set of all random variables representing profits and losses of financial positions and $\mathcal{M}$ the set of all distribution functions. 
\begin{definition} \label{definition:LQ}
Given a function $\Lambda\colon \mathbb{R} \to [\lambda_m, \lambda_M]$, with  $0 < \lambda_m \leq \lambda_M < 1$, we define the lambda quantile of $Y \in \mathcal{X}$, with $F(x) = \mathbb{P}(Y\leq x)$, as follows: 
\begin{equation}\label{def:LQ}
    \rho_{\Lambda}(Y)=\inf\{x: F(x) >\Lambda(x)\},
\end{equation}
and call lambda value at risk, $\Lambda$VaR, the associated risk measure 
$$\operatorname{\Lambda{}VaR}(Y)= -\rho_{\Lambda}(Y).$$
\end{definition}
When the $\Lambda$ function is constant at some $\lambda \in (0,1)$, lambda quantiles align with the traditional definition of quantiles at $\lambda$ confidence level. \Review{As discussed in \cite{bp19}, $\Lambda$-quantiles can be defined in four alternative ways. The definition adopted here} coincides with the original definition introduced in \cite{fmp14}, where lambda value at risk is defined as the negative of the right lambda quantile, due to its interpretation as capital requirement of profits and losses. \Review{The algorithm proposed in this paper remains applicable to all four definitions, as it assumes the uniqueness of the lambda quantile, under which these definitions are equivalent. Ultimately, }the key feature of lambda quantiles is that the confidence level, assigned by the $\Lambda$ function, varies as a function of the asset's losses, allowing for more flexibility and better capturing different tail behaviours. 

\subsection{Essential properties}

Lambda quantiles possess key properties that make them suitable as risk measures. Primarily, they are monotonic increasing, meaning they assign higher risks to greater losses, a fundamental feature that all risk measures should inherently satisfy. Formally:  
$$ \rho_\Lambda(X) \leqslant \rho_\Lambda(Y) \text{ for all } X, Y \in \mathcal{X} \text{ with } X \leqslant Y.$$
Hence, the $\Lambda$VaR is monotone decreasing.

As many other risk measures in finance, lambda quantiles are law invariant since  they assign the same value to random variables with the same distribution, $$
X \sim_d Y \Rightarrow \rho_\Lambda(X)=\rho_\Lambda(Y).
$$
For this reason, we can alternatively define the lambda quantile on distribution functions and set, with a slight abuse of notation, that $\rho_\Lambda(Y)=\rho_\Lambda(F)$.

%CASH SUBADDITIVITY AND STAR SHAPEDENESS MAYBE..(NOT SURE).

Another important property that is particularly important when estimating functionals is the elicitability~\cite{gneiting2011making}. $\Lambda$-quantiles, as singled-valued functionals, are elicitable~\cite{bellini2015elicitable,bpr17} since they admit a \emph{scoring} or \emph{loss} function $S\colon \mathbb{R} \times \mathbb{R} \rightarrow[0,+\infty)$ given by $$S(x, y)=(y-x)^{-}-\int_y^x \Lambda(t) \mathrm{d} t$$ that makes them the unique minimiser of the $\mathbb{E}_F[S(x, Y)]$ for any choice of $F \in \mathcal{M}_\Lambda \subseteq \mathcal{M}$, i.e.
$$
\rho_\Lambda(F)=\arg \min _x \mathbb{E}_F[S(x, Y)] \quad \forall F \in \mathcal{M}_\Lambda
$$
where $\mathcal{M}_\Lambda$ is the domain of elicitability. 
When we interpret $S(x, y)$ as the realized forecasting error between the ex-ante prediction $x =\rho_\Lambda(F) \in \mathbb{R}$ and the ex-post observation of $Y$, $y \in \mathbb{R}$, the importance of this property for any application becomes more evident. Elicitability guarantees the existence of a loss function that is strictly consistent with the statistical functional, allowing statistically meaningful comparisons between different model assumptions on the distribution of $Y$. This facilitates the selection of the model $F$ that minimizes the forecasting error $\mathbb{E}_F[S(\rho(F), Y)]$.

If we take $\Lambda$ decreasing the elicitability of lambda quantiles holds for any distribution function with finite first moment. More generally, the \Review{ elicitability} of lambda quantiles holds under the assumption stated in the following remark.

\begin{remark}[\cite{bpr17}] \label{eliassumpt}
Lambda quantiles $\rho_\Lambda$ are elicitable in $\mathcal{M}_{\Lambda}$ that is the set distribution functions $F \in \mathcal{M}_1$ such that
\begin{equation} \label{eq:eliassumpt}
 \exists \bar{x} \text { s.t. } \forall x <\bar{x}, \; F(x)<\Lambda(x) \text { and } \forall x>\bar{x}, F(x)>\Lambda(x) 
\end{equation}
and $\mathcal{M}_1 \subset \mathcal{M}$ is the subset of distributions with finite first moment. 
\end{remark}
This, alternatively, suggests that somebody can choose a priori a $\Lambda$ function for which this condition holds for any distribution function, which means choosing $\Lambda$ that is crossed only once by any possible $F$ at the level given by the lambda quantile $\bar{x}=q_{\Lambda}(F)$.

\subsection{Risk contribution with lambda quantiles}

When dealing with multiple assets' portfolio, finding assets' contribution to the total risk is a relevant concern. This can be done by computing the derivative of the risk measure with respect to the weight $w_i$ of the asset in the portfolio, $\frac{\partial \rho_{\Lambda}}{\partial w_i}(\bfw)$. 
Using lambda quantiles, this problem has been studied in~\cite{ince:22}, which found the assumptions under which lambda quantile's risk contributions can be provided. 

\begin{remark}[\cite{ince:22}, Assumption 1] \label{rem:Ince}
Let $Y=X(\bfw)$ be the random variable of a portfolio profit and loss with $d$ assets $(X_i)$, $i=1\dots d$ and allocations $\bfw \in \mathbb{R}^d$. Assume the portfolio operator $X(\bfw)$ is monotone increasing with respect to $w_i$ for all $i$.
If the distribution function of $Y$ is continuously partially differentiable in $y$ and $w_i$ for all $i$ and has density function $\varphi$ such that $$
\varphi\left(\rho_{\Lambda}(\bfw)\right)>\Lambda^{\prime}\left(\rho_{\Lambda}(\bfw)\right)
$$
for a given continuously differentiable $\Lambda$, lambda quantiles admit risk contributions.

If this assumption holds, the risk contributions of the $i$-th asset to the total portfolio risk, $\rho_\Lambda(\bfw)$, are given by
\begin{equation}\label{eq:LQgrad}
\frac{\partial \rho_{\Lambda}}{\partial w_i}(\bfw)=\frac{\varphi \left(\rho_{\Lambda}(\bfw)\right)}{\varphi\left(\rho_{\Lambda}(\bfw)\right)-\Lambda^{\prime}\left(\rho_{\Lambda}(\bfw)\right)} \mathbb{E}\left[\partial_{w_i} X(\bfw) \mid X(\bfw)=\rho_{\Lambda}(\bfw)\right].
\end{equation}
\end{remark}

\subsection{Applications of lambda quantiles to risk management}

One of the earliest applications of lambda quantiles in risk management was developed in \cite{hmp18}. Their research investigates the role of the $\Lambda$ function and explores its calibration to enhance the traditional value at risk measure. Their primary insight is to view $\Lambda$ as a proxy for the tail behavior of the market portfolio distribution. This approach enables $\Lambda$VaR to analyze how different assets respond to certain market conditions by identifying quantiles across multiple confidence levels. Additionally, they introduce a dynamic component by regularly updating $\Lambda$, allowing the quantile detected at a certain confidence level to adapt to changing market conditions. 

To calculate $\Lambda$VaR, the authors suggested various models, including one where $\Lambda$ is determined through linear interpolation of $n$ points. If an increasing $\Lambda$ is considered these points lie between the lower point $(x_m,\lambda_m)$ and the upper point $(x_M,\lambda_M)$, and $\Lambda$ maintains a constant value equal to $\lambda_m$ to the left of $x_m$ and $\lambda_M$ to the right of $x_M$. The $x_n$ are computed based on quantiles of the distributions of selected market benchmarks, with $x_m$ chosen as the minimum ever realized return. Compared with a 1\% value at risk, $\lambda_M$ could be increased up to 1.5\% while maintaining good backtesting results. An empirical analysis was conducted using equities in various European markets, comparing different distribution assumptions, including historical simulations, Monte Carlo Normal, Monte Carlo Student-t, and GARCH(1,1) with t innovations. As pointed out in \cite{corbetta2018backtesting}, the number of simulations $M$ impacts the choice of the minimum value $\lambda_m$, which should be set to a value higher than $1/M$ to ensure that $F$ starts lower than $\Lambda$, thus avoiding the degenerate situation where the lambda quantile detects only the confidence level $\lambda_m$, which might happen when $\Lambda$ is not decreasing.

The previous empirical application on lambda quantiles were \Review{expanded} in~\cite{corbetta2018backtesting} by introducing a comprehensive backtesting framework. In particular, they construct three different hypothesis tests based on violations not identically distributed, since for lambda quantiles the probability $\mathbb{P}(I_t=1)=\Lambda(\rho_{\Lambda,t})$ changes at each time $t$, but assumed independent since $\Lambda$ is recalibrated at each time $t$ on past information. Leveraging this assumption, the authors employed results from probability theory concerning random variables that are independent but not identically distributed to construct test statistics. Among them, the simplest test uses the sum of violations as a test statistic, which follows the Poisson Binomial distribution and allows for a framework very similar to VaR. In their empirical application, they computed lambda quantiles using additional distribution assumptions, including bootstrapped historical simulations and an extreme value theory approach with the generalized Pareto.

Lambda quantiles in a portfolio of equity options are used in~\cite{ince:22}, computing risk contributions and Euler allocations. They reduced the calibration points by employing an exponential $\Lambda$ and ensuring it passes through $(x_m,\lambda_m)$ and $(x_M,\lambda_M)$ while maintaining constancy elsewhere, consistent with prior applications. Given the necessity for the return distribution to be differentiable to meet the risk contributions assumption, they adopted a kernel density estimation approach applied to historical values. 

However, these approaches do not provide any guidance on how to actually compute the lambda quantile, but just suggest in some cases the use of a built-in solver function available in any programming language. The following section, we show the limitation of this approach and provide an efficient algorithm to compute $\Lambda$-quantiles under quite general conditions.

\section{Globally Convergent Newton-type Method for $\Lambda$-Quantiles}\label{sec:NewtonBis}

We start this section with a motivational example that highlights the need of a dedicated method to solve lambda quantiles. Consider an asset with profit and loss $X \in \mathcal{X}$ and distribution function $F$. We assume that the lambda quantile of $X$, as defined in \eqref{def:LQ}, is \Review{the unique sign-change of $F-\Lambda$}, essentially requiring that $F$ crosses $\Lambda$ at a single point. Further details can be found in Assumption \ref{assumption1} in Section \ref{sec:convergence}. For an initial investigation, we consider the simplest case that both $F$ and $\Lambda$ are continuous. Computing the lambda quantile in this case reduces to solving the non-linear equation:
\begin{equation}\label{eq:lq}
f(x):=F(x) - \Lambda(x)=0.
\end{equation}
Notice that as $\lim_{x\rightarrow -\infty} F(x) = 0$ and  $\lim_{x\rightarrow +\infty} F(x) = 1$, there exist $\xmin, \xmax \in \R$, such that $F(\xmin) \leq \lambda_m$ and $F(\xmax)\geq\lambda_M$, for $0<\lambda_m < \lambda_M < 1$. This ensures $f(x) \leq 0$ for $x\leq \xmin$ and $f(x) \geq 0$ for $x\geq \xmax$. For instance, if $F^{-1}$ exists, $\xmin = F^{-1}(\lambda_m)$ and  $\xmax = F^{-1}(\lambda_M)$ fulfil these requirements.

Conventional methods for solving lambda quantiles involve using general-purpose equation solvers like Matlab's \texttt{fsolve} or the \texttt{fsolve} function from Python's scipy package, which are both based on a Newton method. However, as the following example demonstrates, some of these algorithms often lack the necessary robustness, as they may fail to converge even in relatively straightforward scenarios.

\begin{figure}
\begin{center}
\includegraphics[width=\textwidth]{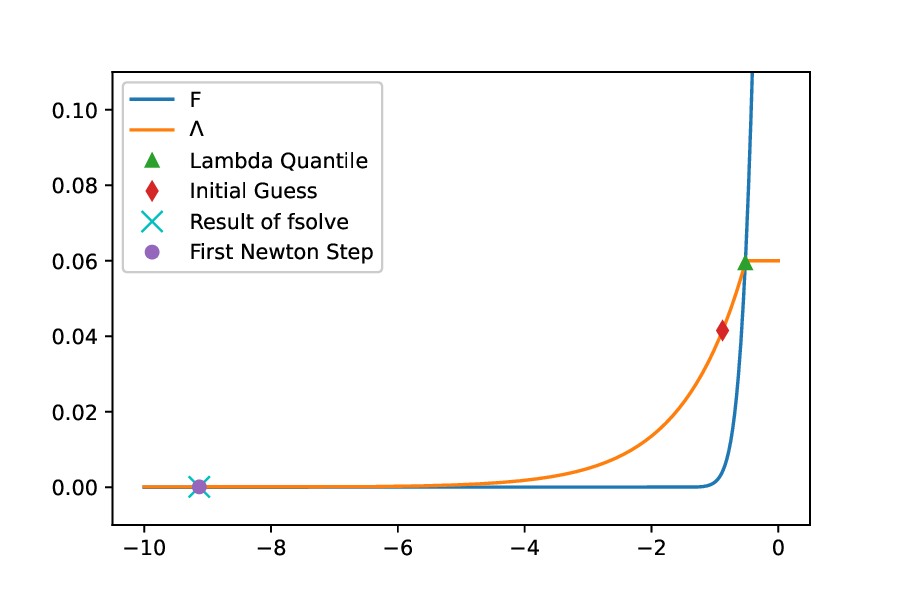}
\end{center}
\caption{Divergence of iteration using a general purpose equation solver. }
\label{fig:first_example}
\end{figure}

\begin{example}\label{ex:example1}
Let $X\sim\mathcal{N}(0,1/3^2)$ and $\Lambda$ be a continuous function given by
\[\Lambda(x) = 
\begin{cases}
0.1 e^x, \quad &   \log(10 \lambda_m)\leq x < \log(10\lambda_M),\\
\lambda_m, \quad &  x  < \log(10 \lambda_m),\\
\lambda_M, \quad &   x\geq \log(10\lambda_M),\\
\end{cases}
\]
with $\lambda_m = 10^{-4}$ and $\lambda_M = 0.06$. Based on a precise numerical evaluation, the lambda quantile of $X$ should be approximately $\bar x=-0.519755\ldots$. To evaluate the performance of general-purpose solvers, we begin by selecting an appropriate starting point - since such solvers typically require one. Given that $\Lambda(x)\in [\lambda_m, \lambda_M]$, we know a-priori that the lambda quantile $\bar x$ will lie within the interval $[q_{\lambda_m}, q_{\lambda_M}]$. Thus, we choose the midpoint of this interval $x_0 = (q_{\lambda_m} + q_{\lambda_M})/2 \approx -0.878965$ as a natural and reasonable initial guess for the iterative procedure. However, an experiment with fsolve from scipy.optimise (using scipy version 1.11.0 and python 3.9.7) shows that this approach does not converge to the true lambda-quantile. Instead, the process terminates with the final iterate $\tilde x \approx -9.130751$, yielding an absolute error of $-8.610995$, with a residual value of $F(\tilde x) \approx 10^{-4}$. This result is shown in Figure~\ref{fig:first_example}, which highlights the discrepancy between the true solution and the one produced by the general-purpose solver. The failure is attributable to the underlying method used by \texttt{fsolve}, which relies on the Newton's method with finite-difference approximation of the first derivative. A direct application of Newton’s method using the same initial guess replicates this behavior: the first iteration produces a result nearly identical to that returned by \texttt{fsolve}, the second iteration explodes to a value of the order of $10^{158}$ and subsequent iterations surpass the upper limit of double-precision floating-number representation. Similar to the Python's default solver, also the Matlab's \texttt{fsolve} (tested in R2022b)is unable to locate the correct lambda quantile for this example, returning a value of $-4.8790$.

\end{example}

\subsection{Newton-type algorithm for computing lambda quantiles}\label{sec:NewotonBisAlg}
The convergence issues highlighted in the preceding example emphasize the necessity of investigating more robust iterative solvers for lambda quantiles. The state of the art for root-finding in the context of smooth functions mainly involves Newton methods~\cite{deuflhard:11}, which are widely praised for their fast local quadratic convergence. The Newton method is an iterative technique that relies on linearizing the function $f$. Starting from an initial value $x_0$, the method updates this value according to the formula:
\[
x_1 = x_0 - \frac{f(x_0)}{f'(x_0)}.
\]
While Newton methods show local convergence, they often struggle with global convergence, as seen in Example~\ref{ex:example1}. To address this limitation, several globalisation techniques have been developed. Among the more popular are methods that reduce large step-sizes, for example through damping~\cite{nesterov:18} or a trust-region~\cite{sorensen:82trustregionnewton}. A discussion of these methods is provided in \cite{deuflhard:11}, where, however, it is noted that ``a guaranteed convergence will only occur, if additional global structure on F can be exploited" \cite[Chapter 3]{deuflhard:11}, for example convexity. For instance, in cases involving one-dimensional problems, the knowledge of the function's sign can significantly enhance the reliability of root-finding algorithms, ensuring convergence. For continuous functions, the intermediate value theorem ensures that there is at least one root between points where the function values have opposite signs, forming the basis for bisection methods~\cite[Chapter 6.2.1]{quarteroni:07}.

Let us now focus on our specific objective of calculating the lambda quantile. As discussed above, we know that $F(x)<\Lambda(x)$ for $x\rightarrow-\infty$ and $F(x) > \Lambda(x)$ for $x\rightarrow \infty$. This observation justifies the use of a bisection method. However, bisection methods converge more slowly compared to Newton methods, so we suggest a hybrid approach. By combining these techniques, we develop a globally convergent method that also achieves local quadratic convergence for differentiable functions.

\begin{figure}
\includegraphics[width=.5\textwidth]{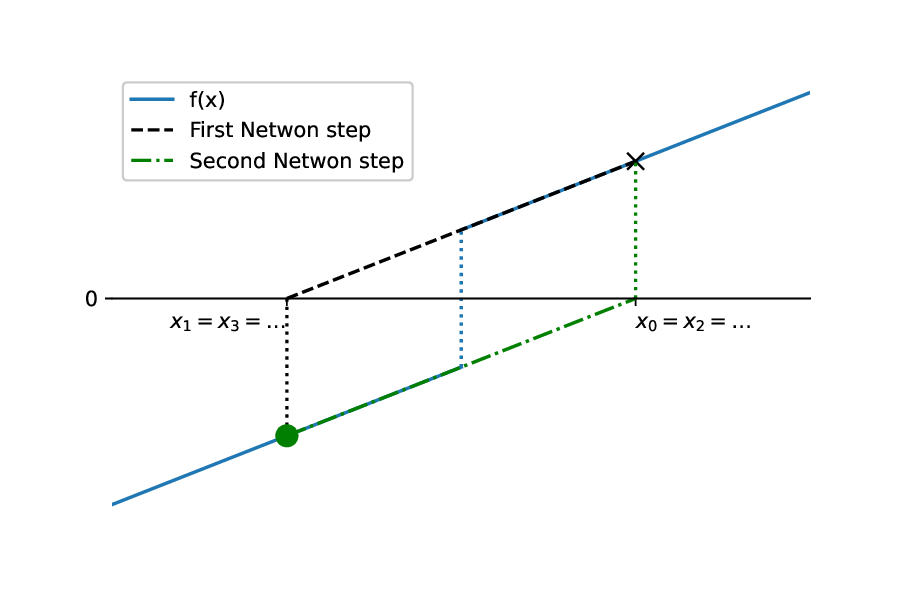}\hfill
\includegraphics[width=.5\textwidth]{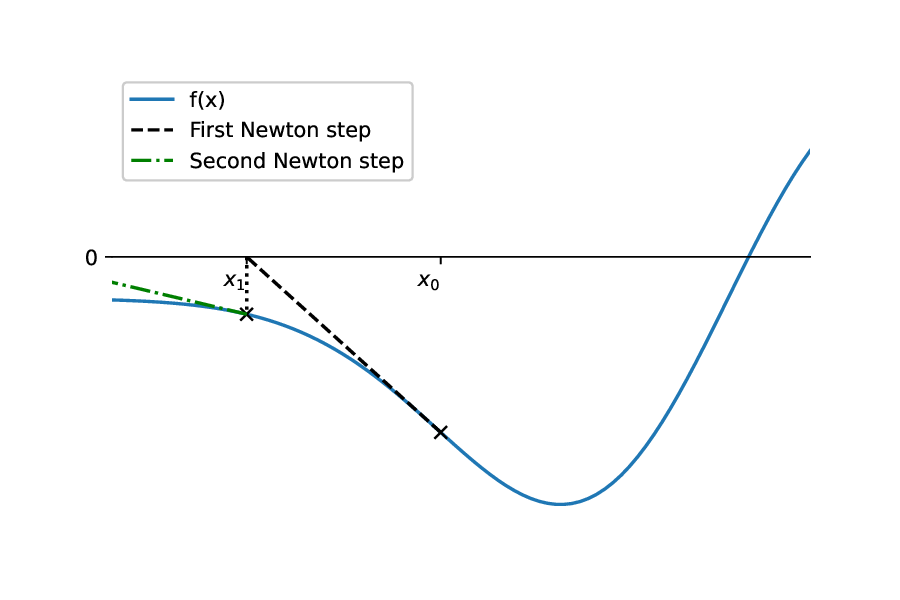}
\caption{Two main Newton failure modes: Oscillation (left) and divergence (right).}
\label{fig:newton_failures}
\end{figure}

To determine when to perform a Newton step versus a bisection step, we need to understand the situations in which the Newton method fails to converge. As presented in Figure~\ref{fig:newton_failures}, two main failure modes can occur: oscillation and divergence. Oscillation arises when, after two or more Newton steps, the process returns to the original iterate. In the case of divergence, the sequence of Newton iterates becomes unbounded, failing to converge to the root. For instance, in Example~\ref{ex:example1}, the second scenario, divergence, is responsible for the inability to correctly solve the equation. 
 
Combining the Newton method with a bisection approach effectively avoids the second failure mode. Specifically, we prevent divergence by ensuring all iterates remain within a finite interval. Therefore, if we observe that the updated estimate satisfies
\[
x_i - \frac{f(x_i)}{f'(x_i)} \not\in [x_\mathrm{l}^i, x_\mathrm{r}^i],
\]
we reject the Newton step and perform a bisection step, maintaining the iterates within the prescribed bounds.

Nevertheless, oscillation can still occur. A naive approach would be to ensure that the bisection interval keeps decreasing by taking bisection steps. As the length of the bisection interval approaches zero, the iterates naturally converge, eliminating the possibility of oscillation. However, the slow convergence of the bisection method would severely slow down the algorithm. For this reason, we choose a bisection step only if we detect potential oscillations by comparing the step-size with the length of the interval. These can occur when the step-size is large compared to the length of the interval. For example, consider $x_i=x_\mathrm{l}^i$, we reject the Newton step and perform a bisection step if the new iterate $x_i + \frac{f(x_i)}{f'(x_i)}$ is close to $x_\mathrm{r}$.

Combining both conditions into a single one we only accept the Newton iteration if
\[
x_i - \frac{f(x_i)}{f'(x_i)}  \in \left(x_\mathrm{l}^i+\delta \left|\frac{f(x_i)}{f'(x_i)} \right|, x_\mathrm{r}^i - \delta \left|\frac{f(x_i)}{f'(x_i)} \right|\right),
\]
for some $0<\delta < 1/2$. The whole algorithm, called\ouralgorithm{}, is given in Algorithm~\ref{alg:newton}.

Note that the idea of combining a slow but globally convergent bisection method with a faster converging algorithm goes back decades, with Brent's method~\cite{brent:71} being a famous example. Brent's method combines the bisection method with inverse quadratic interpolation and does not require evaluations of the derivatives. In contrast, the algorithm proposed here applies Newton steps, which are based on the function's derivatives. In this specific application, evaluating the derivatives is more efficient because the equation solving for the lambda quantiles is based on the    distribution function, whose derivative, i.e. the density function, is often computationally easier to evaluate. 

\begin{algorithm}[H] \small
   \LinesNumbered

 \KwData{Input: Function handles for $f$ and $f'$, $\xmin, \xmax$. \\Parameters $0< \delta < 0.5$, $N_{\max{}} \in\N_{>0}$, $\varepsilon >0$ (e.g. $\delta  = 0.01$, $N_{\max{}}= 100$, $\varepsilon= 10^{-8}$) }
 \KwResult{Lambda quantile $\bar x$ }
$x_0 \leftarrow (\xmin + \xmax)/2$, $x_l \leftarrow \xmin$, $x_r \leftarrow \xmax$\;
\uIf{$f(x_l)<\varepsilon$}{
\Return{$x_l$}\;
}

\ElseIf{$f(x_r)<\varepsilon$}{
\Return{$x_r$}\;
}
\For{$i \leftarrow 1$ \KwTo $N_{\max{}}$}{
  $f_0 \leftarrow f(x_0)$\;
  \uIf{$\operatorname{abs}(f_0) < \varepsilon$}{
   \Return{$x_0$}\;
   }
   \uElseIf{$f_0 < 0$}{
   $x_l \leftarrow x_0$\;
  }
  \Else{
  	$x_r \leftarrow x_0$\;
 }
  \If{$x_r-x_l < \varepsilon$}{
 \Return{$x_0$}
 }
\tcp{Next iteration computed from here}
 $df_0\leftarrow f'(x_0)$\;
 $dx \leftarrow -f_0/df_0$ \tcp*{ if $df_0=0$, we set $dx$ to be infinite} 
 $x_0 \leftarrow x_0 + dx$\; \label{line:newton}
 \If{$x_0 \geq x_r - \delta \operatorname{abs}(dx)$ or $x_0 \leq x_l + \delta \operatorname{abs}(dx)$\label{line:bisection_condition}}{
 	$x_0 \leftarrow (x_l + x_r)/2$\; \label{line:bisection}
 	}
 }
 \Return{No convergence within $N_{\max{}}$ steps}\;
 \caption{\ouralgorithm: globally convergent Newton method for $\Lambda$-quantiles}\label{alg:newton}
\end{algorithm}

\subsection{Convergence and rate of convergence} \label{sec:convergence}

Here, we present the assumptions under which the proposed\ouralgorithm{} algorithm converges. We demonstrate that convergence is ensured under weak assumptions (Assumption \ref{assumption1}), while slightly stronger regularity assumptions are necessary to establish local second-order convergence (Assumption \ref{assumption3}). 

\begin{assumption}
\label{assumption1}
Assume that \Review{for $X\sim F$}:
\begin{itemize}[left=.7em]
    \item[(i)] 
    \Review{There is a unique sign-change of the function $F(x) - \Lambda(x)$}, i.e. there exists $\bar x \in \R$ with 
\[F(x)<\Lambda(x) \quad x<\bar x \text{ and } F(x) > \Lambda(x) \quad x > \bar x.
\]

\item[(ii)] The function $f(x) = F(x) - \Lambda(x)$ is a càdlàg, i.e. right continuous with left limits on $\R$ with a càdlàg right-derivative, denoted $f'$. Furthermore, the left-sided limit can only be zero at the lambda quantile, i.e. $f(x-) = 0 \Rightarrow x=\bar x$.
\end{itemize}
\end{assumption}
 This assumption ensures global convergence while allowing for jumps in both $F$ and $\Lambda$.
Note that \Review{$\bar x$ in Assumption \ref{assumption1} (i) is equal to  the lambda quantile of $X$.}  Additionally, Assumption \ref{assumption1} (i) aligns with condition \eqref{eq:eliassumpt} in Remark~\ref{eliassumpt}, preserving the elicitability of the lambda quantile computed using the\ouralgorithm{} algorithm.

To recover quadratic convergence of the Newton method, we need an additional regularity condition, as shown by the following assumption. 
\begin{assumption}
\label{assumption3}
In addition to Assumption~\ref{assumption1}, let $f\in C^2$ in an interval around $\bar x$ and $f'(\bar x)\neq 0$.
\end{assumption}

The following theorems demonstrate the convergence of our algorithm. When analyzing convergence, we consider the case of no early stopping, i.e. formally $N_{\max{}} = \infty$ and $\varepsilon=0$. If the iteration successfully concludes with $x_n = \bar x$, we consider $x_i = \bar x$ for all $i>n$.
The proofs of these theorems are provided in Appendix~\ref{sec:appendix_proofs}.

\begin{theorem}\label{thm:convergence}
Under Assumption~\ref{assumption1}, the sequence of $x_0$ created by Algorithm~\ref{alg:newton} (without early stopping) converges towards the unique lambda quantile $\bar x$. 
For sufficiently large $N_{\max{}}$, the algorithm stops when $x_0$ fulfils
\[|f(x_0)| < \varepsilon \text{ or } |x_0 - \bar x| < \varepsilon.\]
\end{theorem}

Next, we show that under reasonable additional assumptions our algorithm preserves local quadratic convergence.

\begin{theorem}\label{thm:quadratic_convergence}
Under Assumption~\ref{assumption3}, the final convergence rate is quadratic, i.e. there exists $n\in\N$ and $C>0$, such that for all $i>n$:
\[
|e_{i+1}| \leq C |e_i|^2
\] 
with $e_i$ being the absolute error of the $i$-th iterate, $e_i=|x_i-\bar{x}|$. 
\end{theorem}

\begin{remark}\label{rem:several_roots}
In Theorem~\ref{thm:convergence} we have shown global convergence of the iteration to a generalised root of the equation $F(x) = \Lambda(x)$. Under Assumption~\ref{assumption1} this is the unique generalised root and hence the lambda quantile in \eqref{def:LQ}.

If the Assumptions are not fulfilled and there are several roots, Theorem~\ref{thm:convergence} only guarantee convergence to one of them, not necessarily the lambda quantile. However, we note that by construction the algorithm only converges to a root where $F(x-\varepsilon) \leq \Lambda(x-\varepsilon)$ and $F(x+\varepsilon) \geq \Lambda (x+\varepsilon)$. In typical cases, this excludes about half of the roots, for example when there are   $2n+1$ single roots, the algorithm converges to one of $n+1$, with the smallest one being the lambda quantile.

In these cases, convergence to the lambda quantile is not guaranteed due to the local nature of numerical schemes. Ideally, properties of the function $F(x) - \Lambda(x)$, including their derivative or curvature, can be used to determine a closer range in which the lambda quantile lies. If we can determine an interval with $\bar x \in [x_\mathrm{min}, x_\mathrm{max}]$, where $F(x_\mathrm{min}) < \Lambda(x_\mathrm{min})$ and $F(x_\mathrm{max}) > \Lambda(x_\mathrm{max})$ where only one root exists (e.g. due to $F'(x) > \Lambda'(x)$ for $x \in [x_\mathrm{min}, x_\mathrm{max}]$), the presented algorithm converges towards the lambda quantile.

If no prior knowledge on $F-\Lambda$ can be exploited, a Monte-Carlo approach can be considered, sampling $F-\Lambda$ on a fine grid to locate sign changes. 
An alternative approach to locating all roots of an equation is given by the interval Newton method, based on interval analysis~\cite{moore:09}. Refer to Section \ref{sec:interval} for a preliminary investigation of this situation.
\end{remark}

\subsection{Practical examples}
In this section, we evaluate the performance of the proposed method in situations of practical relevance. This includes challenging cases, such as bimodal distributions and discontinuities. In all examples, we report the number of steps required and whether Newton or bisection steps were taken. 

Through this section, we use piece-wise exponential lambda functions of the form
\begin{equation}
\label{eq:exponential_lambda}
\textstyle
\Lambda(x) = 
\begin{cases}
\lambda_m, \quad &x<x_m\\
\beta e^{\alpha x}, \quad & x_m \leq x < x_M,\\\lambda_M, \quad & x\geq x_M,\\
\end{cases}
\end{equation}
where the parameters considered are specified in each example. We consider normal, Student-t and a double Weibull distribution to demonstrate robustness in a wide range of situations. 

\subsubsection{Normal distribution: Revisiting Example~\ref{ex:example1}}
As a first example, we revisit the motivating example, where we have seen a failure of Python's standard solver and the classical Newton method. 

%We note that the lambda function of Example~\ref{ex:example1} is of the form~\eqref{eq:exponential_lambda} for $\alpha = 0.1$, $\beta = 1$, $\lambda_m = 10^{-4}$, $\lambda_M=0.06$, $x_m = \log(\lambda_m/\alpha)$ and $x_M = \log(\lambda_M/\alpha)$. %% I don't think this adds anything.

\begin{figure}
    \centering
    \includegraphics[width=.4\textwidth]{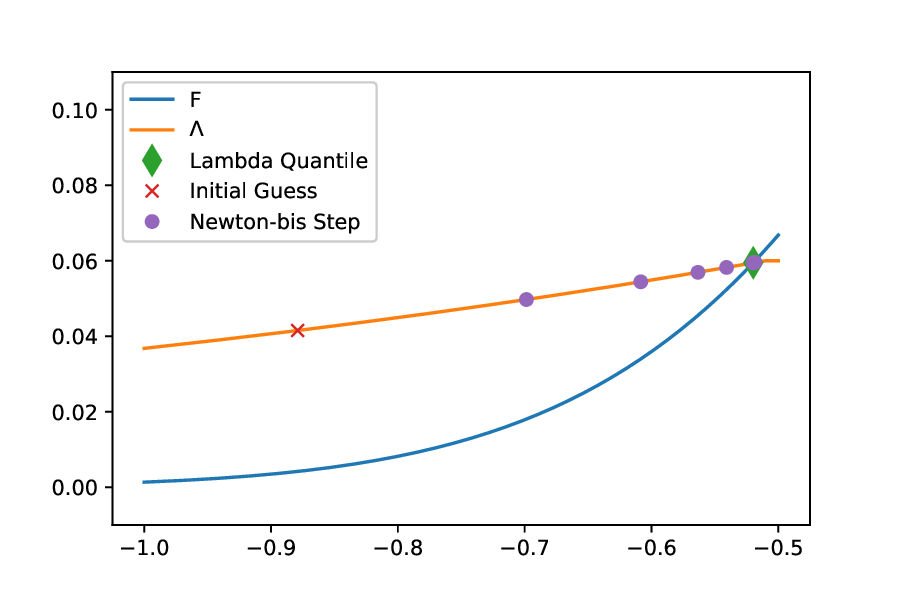}\hspace{1em}
    \includegraphics[width=.4\textwidth]{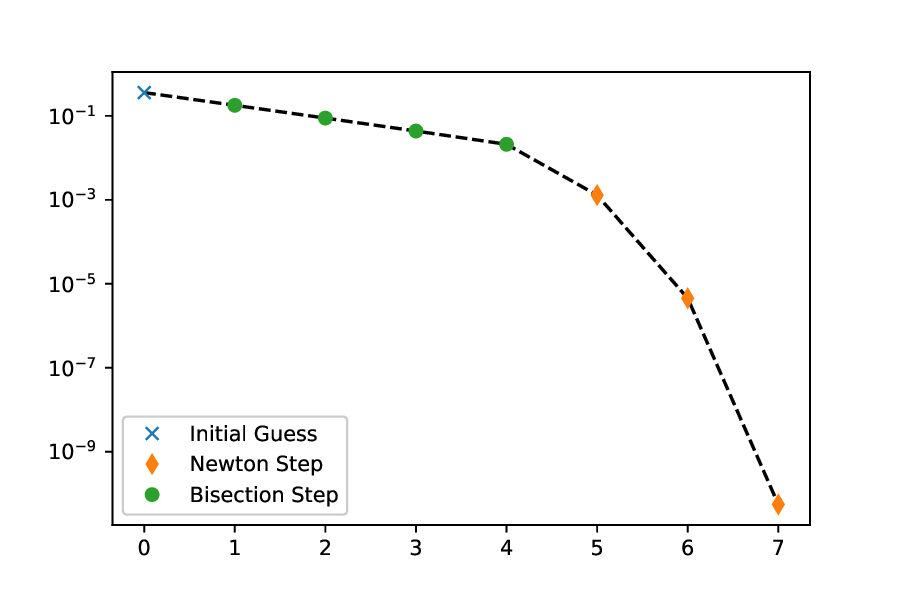}
    \caption{Convergence of the\ouralgorithm{} algorithm for Example~\ref{ex:example1}. Left: Plot of the iterates with $F$ and $\Lambda$ shown as well. Right: Convergence of the iterations with an indication of when Newton or bisection steps were taken. The error has been evaluated by comparison to an iteration with a smaller tolerance.}
    \label{fig:first_example_newton_bis}
\end{figure}

In Figure~\ref{fig:first_example_newton_bis} we see that the\ouralgorithm{} algorithm could overcome the issues faced by other solvers. We observe convergence within just a couple of steps. As expected from the divergence of the pure Newton method, the first steps were taken using the bisection algorithm. This guarantees stability of the iterates. After four bisection steps, we have entered the region of local convergence and just three Newton steps are sufficient to return a solution up to an error of less than $10^{-9}$.

\subsubsection{Student-t distribution}
\begin{figure}
    \centering
    \includegraphics[width=.4\textwidth]{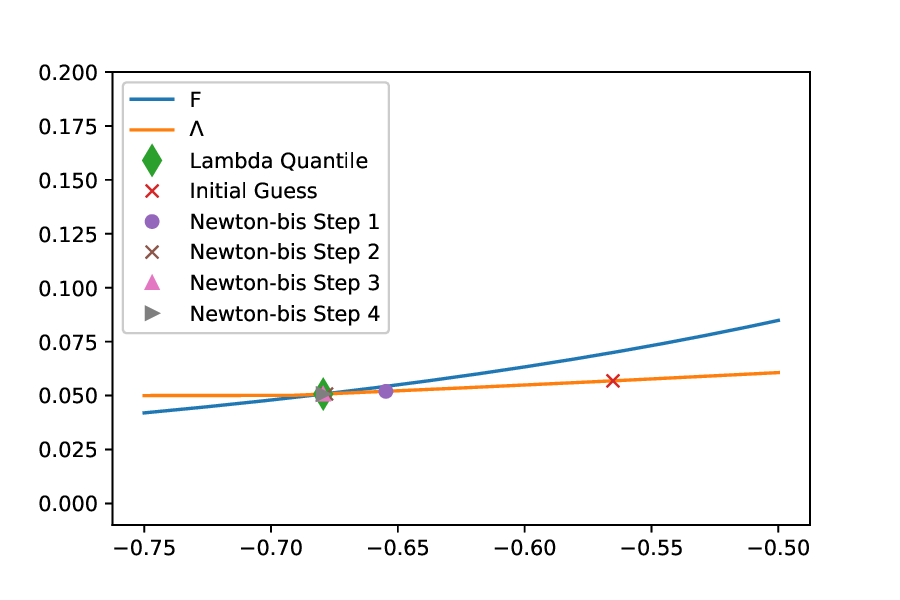}\hspace{1em}
    \includegraphics[width=.4\textwidth]{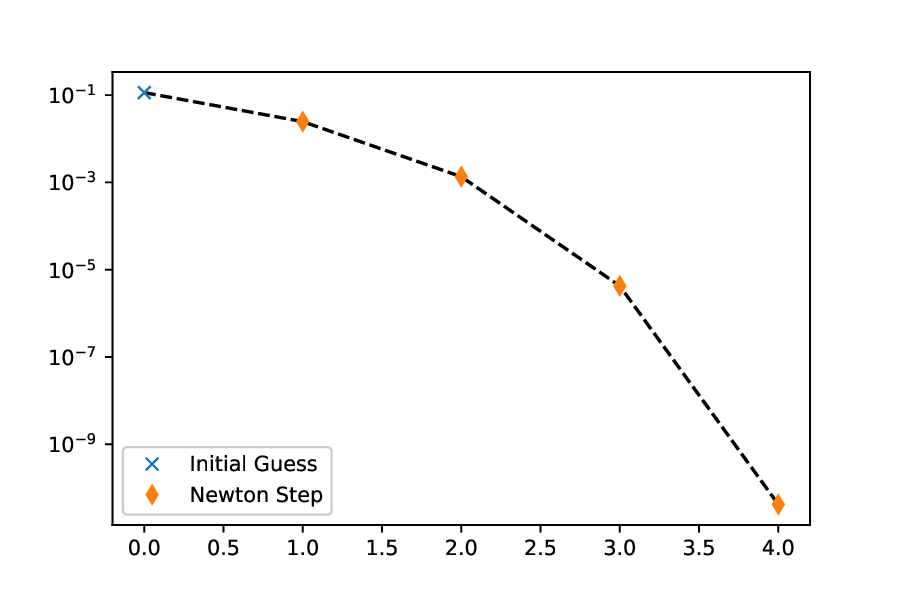}
    \caption{Convergence of the\ouralgorithm{} algorithm for a Student-t example. Left: Plot of the iterates with $F$ and $\Lambda$ shown as well. Right: Convergence of the iterations where only Newton steps were taken. The error has been evaluated by comparison to an iteration with a smaller tolerance. }
    \label{fig:example_t_distribution}
\end{figure}

We consider an example using the Student-t distribution. In particular, we assume $X\sim t_\nu(\mu, 
\sigma^2)$ with $\nu=3$, $\mu=0.1$, $\sigma=1/3$. For a lambda function with the structure~\eqref{eq:exponential_lambda}, we set the parameters as follows: 
$\alpha = 0.1$, $\beta = 1$, $\lambda_m = 0.05$, $\lambda_M=0.1$, $x_m = \log(\lambda_m/\alpha)$ and $x_M = \log(\lambda_M/\alpha)$. 
Results are shown in Figure~\ref{fig:example_t_distribution}. In this example, Newton steps were admissible from the start and within just four steps we have identified the solution up to an error of $10^{-9}$.

\subsubsection{Bimodal distribution}

Bimodal distributions can be challenging for evaluation of lambda quantiles as the target function $F-\Lambda$ can be non-monotonic and features local minima. In this chapter, we investigate the performance of the proposed algorithm in these challenging situations. 

To this extend, we consider $X$ distributed as a centered double Weibull distribution~\cite{balakrishnan:85} with $c=5.07$ and lambda of the form~\eqref{eq:exponential_lambda} with $\lambda_m = 0.1, \lambda_M = 0.6, x_m = -3, x_M = 1$. Additionally, the parameters $\alpha = \frac{\log(\lambda_M)-\log(\lambda_m)}{x_M-x_m}$ and $\beta = \lambda_M \left(\lambda_m/\lambda_M\right)^{\frac{x_M}{x_M-x_m}}$ have been set to ensure global continuity of $\Lambda$. 

\begin{figure}
    \centering
    \includegraphics[width=.45\textwidth]{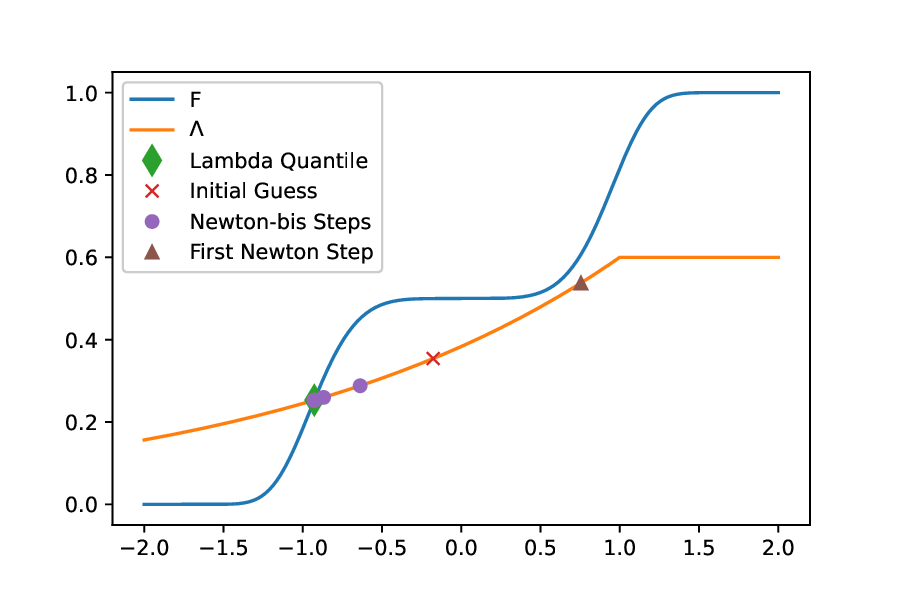}\hspace{1em}
    \includegraphics[width=.45\textwidth]{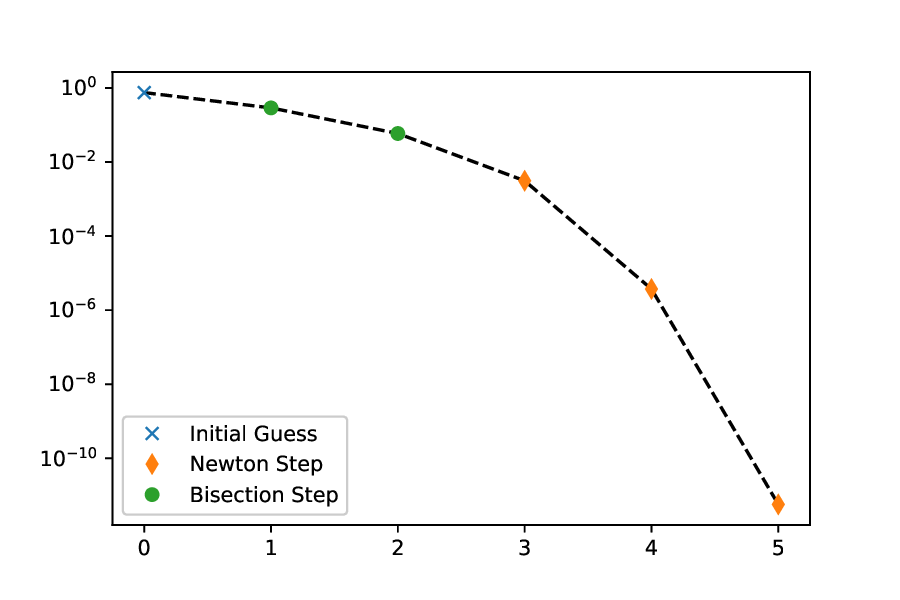}
    \caption{Convergence of the\ouralgorithm{} algorithm for a bimodal double Weibull distribution. Left: Plot of the iterates with $F$ and $\Lambda$ shown as well. For illustration also the first iterate of a pure Newton is shown. Right: Convergence of the iterations with an indication of when Newton or bisection steps were taken. The error has been evaluated by comparison to an iteration with a smaller tolerance. }
    \label{fig:example_dweibull}
\end{figure}

An overview of the problem and the convergence of the algorithm is given in Figure~\ref{fig:example_dweibull}. Notice that the initial value is situated between two modes, which makes a pure Newton algorithm immediately lead towards a local minimum rather than a root. Instead, our algorithm detects this issue and takes two initial bisection steps. These steps are sufficient to reach the area of local convergence. Following this, three more Newton steps reduce the error to below $10^{-10}$. 

\subsubsection{Robustness for discontinuous functions}
Another large class of functions where Newton iterations typically fail to converge are discontinuous functions. However, Theorem~\ref{thm:convergence} shows that our algorithm restores convergence even in case of discontinuities as long as the function is piecewise continuously differentiable. In this section, we explore several practical examples featuring discontinuities that can stem from both the lambda and the distribution function. We examine both cases simultaneously, anticipating consistent performance of the algorithm.

Based on the Student-t distribution, we construct a bimodal distribution with two point masses and an interval of zero probability. For 
\[x_1 = -65,
x_2 = -60,
p_1 = 0.2,
p_2 = 0.6,
\sigma_1 = 1,
\sigma_2 = 2,
\nu_1 = 3,
\nu_2 = 4 
\]
we chose $\mu_1$ and $\mu_2$ such that $F_{\nu_1, \mu_1, \sigma_1}(x_1) = p_1$ and $F_{\nu_2, \mu_2, \sigma_2}(x_2) = p_2$. Here $F_{\nu, \mu, \sigma}$ is the distribution function of the location-scale t distribution.
\[
F(x) = \begin{cases}
    F_{\nu_1, \mu_1, \sigma_1}(x) \quad &  x<x_1\\
    (p_1+p_2)/2 \quad  & x_1\leq x<x_2\\
    F_{\nu_2, \mu_2, \sigma_2}(x) \quad & x\geq x_2
\end{cases}
\]

Two lambda functions $\Lambda_1$ and $\Lambda_2$ are considered of the form~\eqref{eq:exponential_lambda}.

For $\Lambda_1$ we consider $x_m = -80$,
$x_M = -40$,
$\lambda_m = 0.05$,
$\bar\lambda = 0.2$,
$\lambda_M = 0.3$. The parameters $\alpha = \frac{\log(\lambda_M)-\log(\bar\lambda)}{x_M-x_m}$ and $\beta = \lambda_M \left(\bar\lambda/\lambda_M\right)^{\frac{x_M}{x_M-x_m}}$ have been set such that $\Lambda_1$ is continuous at $x_m$ and jumps from $\bar\lambda$ to $\lambda_M$ at $x_M$. 

$\Lambda_2$ features similar properties, but has  a jump  at the same position as  $F$:
This is achieved by selecting $x_m = -78$,
$x_M=-65$,
$\lambda_m = 0.1$,
$\bar \lambda=0.25$ and
$\lambda_M = 0.35$.

\begin{figure}
    \includegraphics[width=\textwidth]{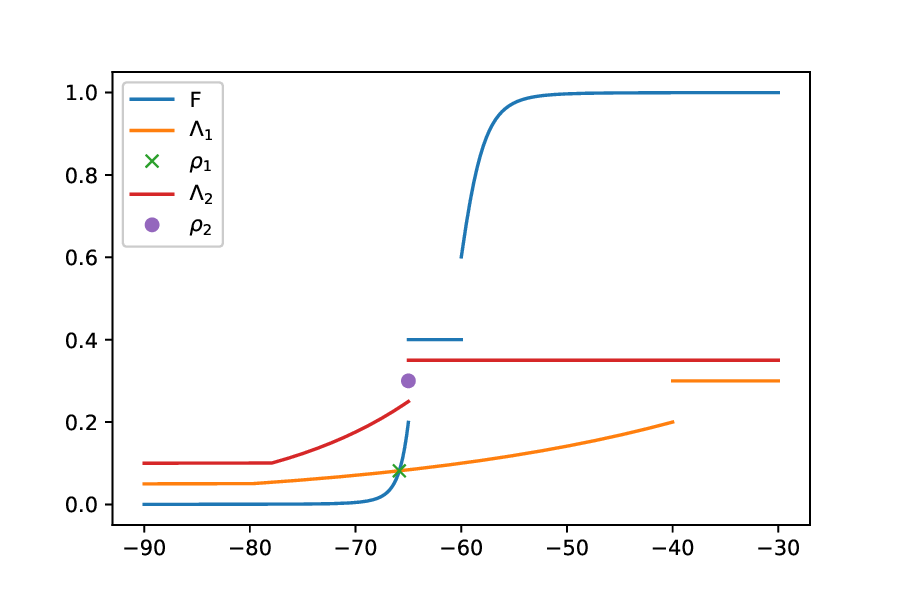}
    \caption{Overview of the distribution and lambda functions in the discontinuous case. }
    \label{fig:discontinuous_example_setting}
\end{figure}

\begin{figure}
    \centering
    \includegraphics[width=.5\textwidth]{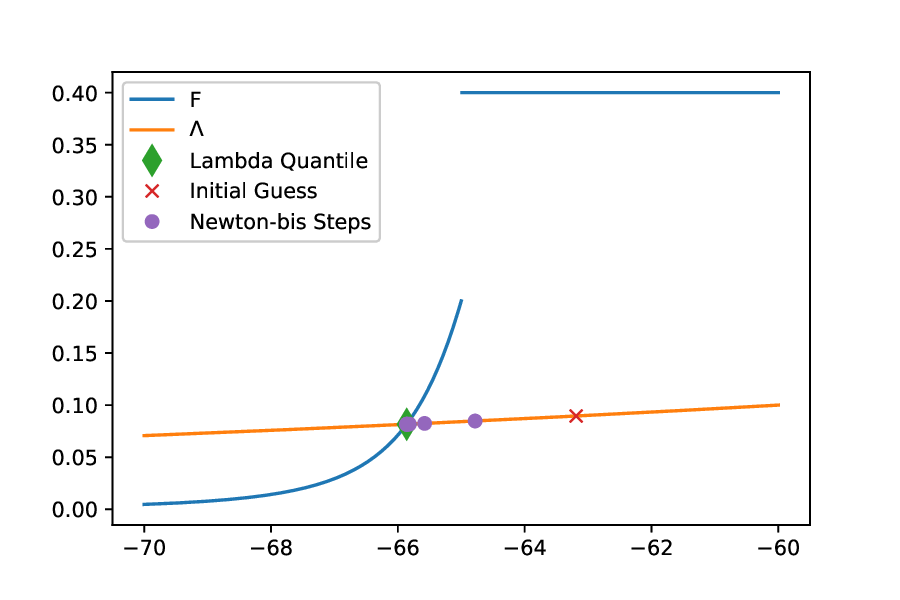}\hfill
    \includegraphics[width=.5\textwidth]{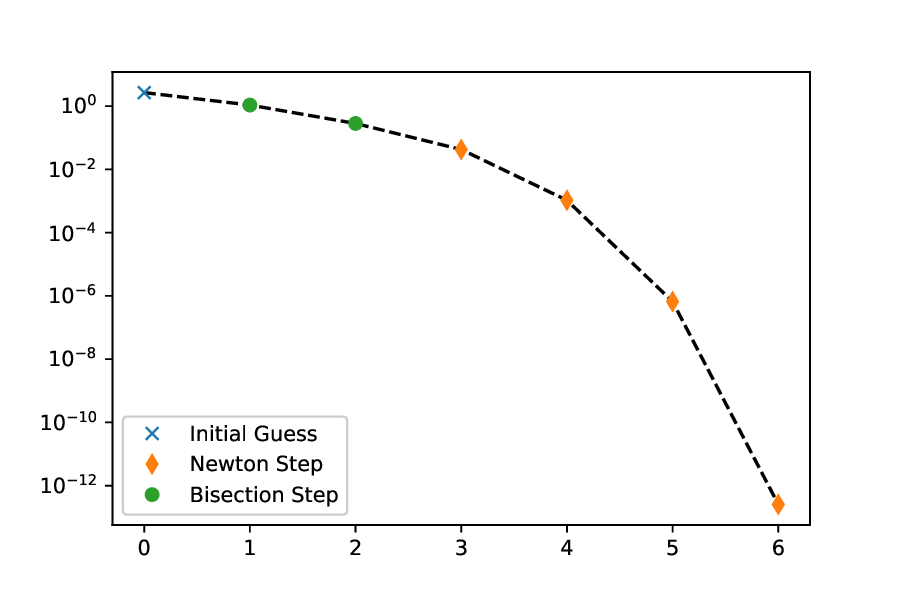}
    \caption{Convergence of the first problem with discontinuous functions. Left: Plot of the iterates with $F$ and $\Lambda$ shown as well. Right: Convergence of the iterations with an indication when Newton or bisection steps were taken.  The error has been evaluated by comparison to an iteration with smaller tolerance. }
    \label{fig:example_discontinuous_1}
\end{figure}

\begin{figure}
    \centering
    \includegraphics[width=.45\textwidth]{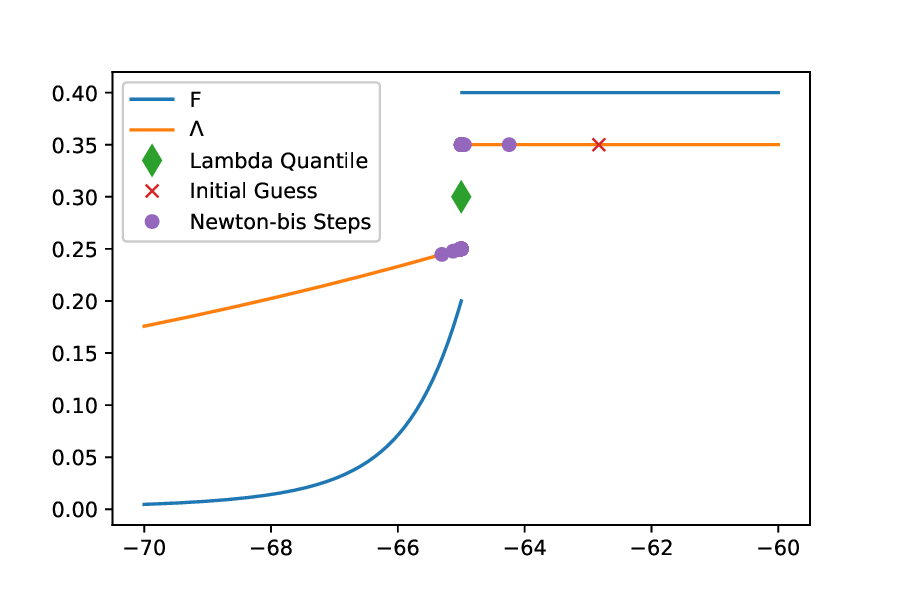}\hspace{1em}
    \includegraphics[width=.45\textwidth]{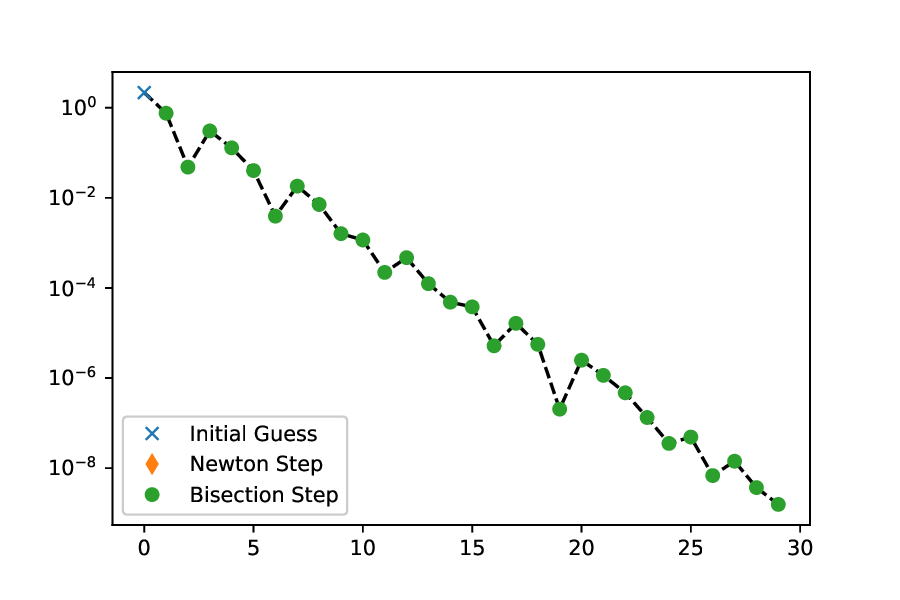}
    \caption{Convergence of the second problem with discontinuous functions. Left: Plot of the iterates with $F$ and $\Lambda$ shown as well. Right: Convergence of the iterations with an indication when Newton or bisection steps were taken.  The error has been evaluated by comparison to an iteration with smaller tolerance. }
    \label{fig:example_discontinuous_2}
\end{figure}

See Figure~\ref{fig:discontinuous_example_setting} for an overview of the distribution and the two lambda functions.    
The results are shown in Figure~\ref{fig:example_discontinuous_1} for the first and Figure~\ref{fig:example_discontinuous_2} for the second case. As expected from our theoretical results, the discontinuities are no obstacle to the convergence of the method. However, we notice a difference in the convergence speed. In the first example in Figure~\ref{fig:example_discontinuous_1}, after just two bisection steps Newton iterates ensure a quick convergence. This is possible as the function is smooth in an area around the solution, guaranteeing quadratic convergence. 
On the other hand, in the second example shown in Figure~\ref{fig:example_discontinuous_2} only bisection steps were taken. This is because the lambda quantile coincides with the discontinuities of $F$ and $\Lambda_2$ and the Assumptions of Theorem~\ref{thm:quadratic_convergence} is not fulfilled. As the Newton method is unable to converge to this solution, the proposed algorithm takes bisection steps. This slows down solution significantly and drastically increases the  number of steps needed to achieve the required accuracy in comparison to the other examples. Nevertheless, this is necessary to achieve convergence of the method.

\subsection{Investigation of lambda quantiles in the presence of more than one root} \label{sec:interval}

In case of several roots, the proposed \ouralgorithm \ algorithm still converges to a root of the equation $F-\Lambda$, but this root may not be the lambda quantile. Instead, by Definition~\ref{definition:LQ}, we need to compute the smallest one of these roots, see also Remark~\ref{rem:several_roots}. Due to the local nature of the iterations, neither a Newton-iteration nor a bisection is able to target a specific root. For this reason, in cases of several roots, we propose to use the paradigm of interval analysis~\cite{moore:09} to obtain an initial estimate of the root. In more details, we aim to find an interval in which the lambda quantile is the single root of the equation $F-\Lambda$. Then the bounds of the interval can be used in the $\Lambda$-Newton-bis algorithm to locate the lambda quantile. 

\begin{figure}
    \centering
    \includegraphics[width=\textwidth]{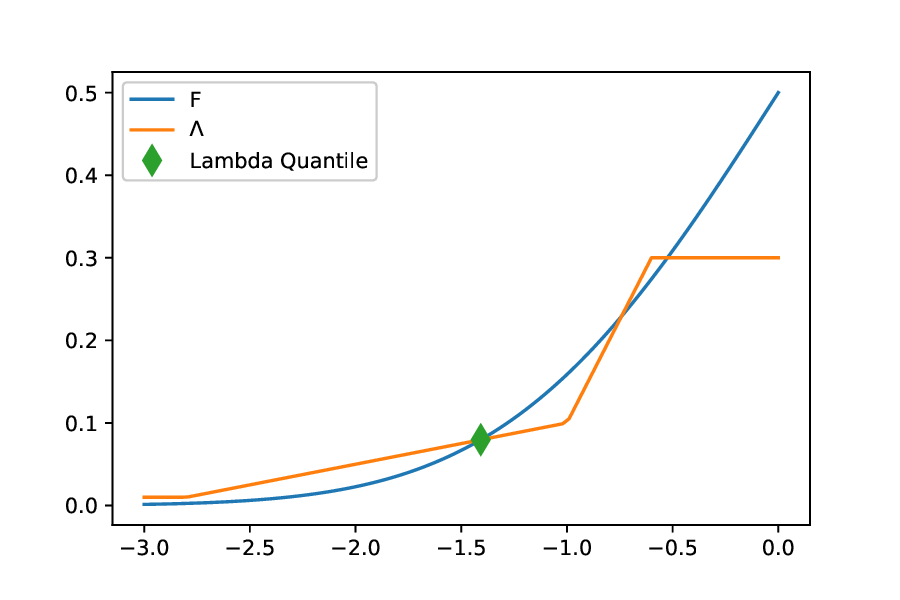}
    \caption{Example with several roots and the location of the lambda quantile.}
    \label{fig:several_roots}
\end{figure}

As an illustrative example, we consider the case of a standard normal distribution $X\sim \mathcal{N}(0,1)$ and a piece-wise linear lambda function
\[\textstyle
\Lambda(x) =
\begin{cases} 
0.01 & \text{if } x < -2.8, \\
0.05x + 0.15 & \text{if } -2.8 \leq x < -1, \\
 0.5x + 0.6   & \text{if } -1 \leq x < -0.6, \\
0.3 & \text{if } x \geq -0.6.
\end{cases}
\]
In Figure~\ref{fig:several_roots}, we note the presence of several roots as well as the lambda quantile. 

Thanks to the monotonicity of $F$ and the suggested  $\Lambda$, we know that for any $x\in[\underline x, \bar x]$, $F(x)\in [F(\underline x), F(\bar x)]$ and $\Lambda(x) \in [\Lambda(\underline x), \Lambda(\bar x)]$. By the rule of subtraction~\cite[Equation 2.16]{moore:09}, we then get that  $F(x) - \Lambda(x) \in [F(\underline x) - \Lambda(\bar x), F(\bar x) - \Lambda(\underline x)]$.
Using this formula to estimate the function range on the interval $[F^{-1}(\lambda_m), F^{-1}(\lambda_M)]$, we find that $F-\Lambda$ does not take any values outside of the value range $[-0.290, 0.266]$. While this is consistent with the presence of a root, it does not contain any information about the possible presence of more than one root. Instead, we have to subdivide the interval to obtain a clearer picture. 

\begin{figure}
    \includegraphics[width=.49\textwidth]{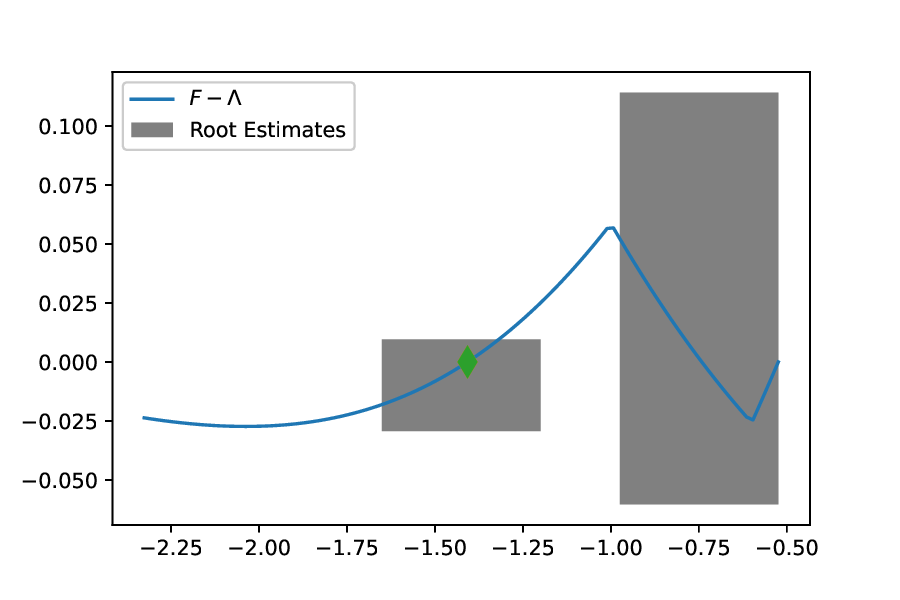}
    \includegraphics[width=.49\textwidth]{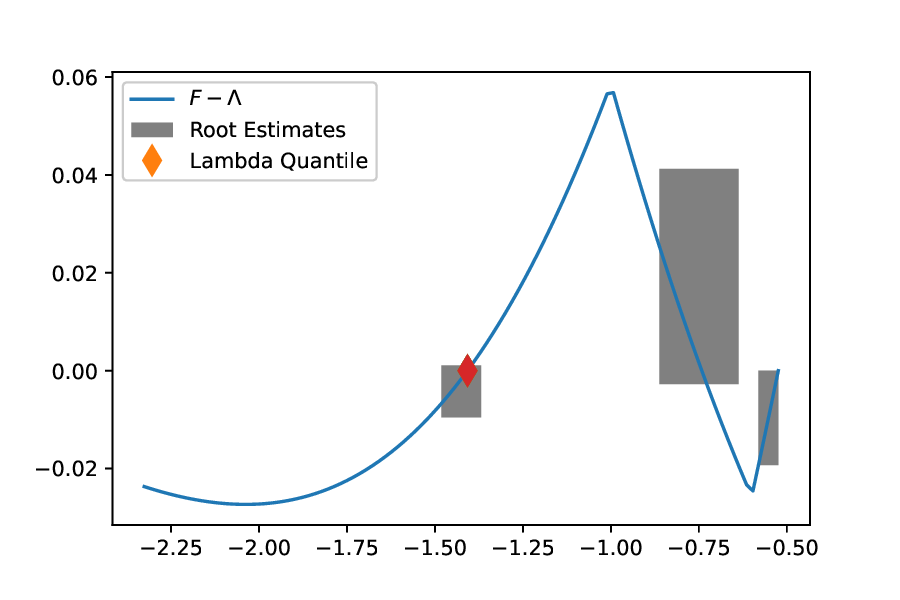}
    \caption{Range estimations using interval analysis with 8 (left) and 32 (right) subdivisions. The grey box represents the estimated interval containing a root and the estimated minimum and maximum of the function over this interval.}
    \label{fig:interval_estimates}
\end{figure}

Already a subdivision into 8 uniform intervals shows that there are at least two roots: at least one in the interval $[-1.65, -1.20]$ and at least one in the interval $[-0.97, -0.52]$. See Figure~\ref{fig:interval_estimates} (left). We note that only 9 evaluations each of $F$ and $\Lambda$ were necessary to obtain this estimate.
We also estimate a function range of $F_\Lambda$ as $[-0.029, 0.010]$ in the interval containing the smallest root. If this range is below our tolerance, we can call the \ouralgorithm \ function to locate the root. 


Should more accuracy be needed, a subdivision into 32 intervals shows that there is at least one root in the interval $[-1.48, -1.37]$ with the function range over this interval being no more than $[-0.0095, 0.0011]$. Two other intervals have been identified to contain a root, but we are interested in the smallest root.  See Figure~\ref{fig:interval_estimates} (right). This estimate is based on 33 evaluations of the two functions. 

We note that while the interval $[-1.48, -1.37]$  could still contain more than one root, the function value between these two roots would not be larger than $0.0011$. If this is below our tolerance, we can locate the root using the Newton-bis method. 
In general, using interval analysis we can identify the number of roots and their estimated location as long as $\max_{x\in[x_i, x_j]} |F(x) - \Lambda(x)| > \mathrm{tol}$ for any two roots $x_i\neq x_j$ and a positive tolerance. 

\subsection{\Review{Application of lambda quantiles to real data}}

\Review{While beyond our primary scope, we briefly discuss the application of lambda quantiles to empirical data. We assume that the function $\Lambda$ is given a priori to focus on computational efficiency of computing the lambda quantile. For $\Lambda$ estimation methods, see \cite{hmp18, corbetta2018backtesting}.}

\Review{The preceding examples assume a parametric approach where the profit and loss follows a specific distribution, such as Normal, Student-$t$, or centered double Weibull. In practice, this involves estimating parameters on in-sample data and computing the lambda quantile out-of-sample using the Newton-Bisection algorithm. However, in many practical settings, one often lacks a known parametric form and must rely on the empirical distribution function. This is standard in historical simulation or Monte Carlo frameworks, where profit and loss realizations are ordered to compute the risk measure. Formally, this shifts the problem from a continuous domain to a discrete one, where $F$ is a piecewise constant step function constructed from $n$  observations $x_1,\ldots, x_n$:$$F_n(x) := \frac{1}{n} \sum_{i=1}^n \mathbf{1}_{\{x_{i} \leq x\}}.$$By this definition, $F_n(x)$ remains constant on each interval $[x_{(i)}, x_{(i+1)})$, where $x_{(1)} \leq \dots \leq x_{(n)}$ denotes the sorted observations. In this discrete setting, the most efficient way to compute the lambda quantile is via sample sorting, as detailed in Algorithm \ref{alg:empirical}.}

\Review{\begin{algorithm}[H]  
   \LinesNumbered

 \KwData{Samples $(x_i)_{i=1,\ldots, n}$, function handle for $\Lambda$. }
 \KwResult{Empirical lambda quantile $\bar x$ }
$({x}_{(i)})_{i=1,\ldots, n} \leftarrow \operatorname{sort}\left((x_i)_{i=1,\ldots, n}\right)$\;

\For{$j \leftarrow 1$ \KwTo $n$}{
  \uIf{$j/n > \Lambda({x}_{(j)})$}{
   \Return{${x}_{(j)}$}\;
   }
}

\caption{Empirical lambda quantiles}
\label{alg:empirical}
\end{algorithm}}

\Review{Notice that, as also highlighted by \cite{corbetta2018backtesting}, one must ensure $n$ is large enough such that $\min(\Lambda)=\lambda_m > 1/n$ to avoid the quantile defaulting to the sample minimum (the worst-case loss). To conclude, the primary computational cost of this algorithm lies in sorting the samples, resulting in an asymptotic complexity of $O(n \log n)$. Since all subsequent steps are performed in linear time, $O(n)$, the overall runtime is dominated by the initial ordering of the data.}

\section{Portfolio Optimisation with Lambda Quantiles}\label{sec:PortLQ}

In this section, we consider an optimal allocation problem based on minimizing risk measured by lambda quantile under some constraints. We propose two alternative methodologies that built on the Netwon-bis algorithm outlined in Section \ref{sec:NewotonBisAlg}, alongside a gradient descent approach incorporating Armijo's line-search for adaptive step-size determination.

\subsection{Theoretical set-up}

A portfolio optimisation problem consists in finding the optimal allocation to assign to each asset based on certain criteria. For a general introduction to non-linear optimisations, see~\cite{nocedal:06} and~\cite{corneujols:18} for applications in finance. A comprehensive review of portfolio optimisation can be found in \cite{kolm201460} and \cite{salo2023fifty}.

We consider $d$ assets, whose returns $X_i, i=1,\ldots,d$ are collected in a random vector $\bfX = (X_i)_i\in\R^d$. Let $\bfmu = (\mu_i)_i\in\R^d$ denote the vector of the expectations, and $\bfSigma\in\R^{d\times d}$ the variance-covariance matrix of the asset's returns. The portfolio allocation is represented by the random vector $\bfw=(w_i)_i \in \mathbb{R}^d$, such that the weights sum to one, $\sum_{i=1}^d w_i=1$, and short selling is not allowed, $w_i \geq 0, i=1,\ldots,d$. For a given allocation $\bfw$, the portfolio's return is represented by the random variable $X(\bfw)=\bfw^\top \bfX= \sum_{i=1}^d w_i X_i\in\R$, with expected return $\mathbb{E}(\bfw^\top X)=\bfw^\top \bfmu=\mu_\bfw$, variance $\sigma^2(\bfw^\top X)=\bfw^\top \bfSigma \bfw=\sigma_\bfw^2$ and lambda quantile $\rho_\Lambda \left (\bfw^\top \bfX \right)$. 

The optimal portfolio $\bfw_{\mathrm{opt}}\in\R^d$ is obtained by minimising risk, quantified as the negative of the lambda quantile, $-\rho_\Lambda\left(\bfw^\top \bfX\right)$, while ensuring a minimum return of $r_\mathrm{min}>0$. Formally, we look for $\bfw_{\mathrm{opt}}$ that solves: 
%\begin{align*}
%w_{\mathrm{opt}} = \operatorname{arg\,max}_{w\in\mathbb{R}^d}~~  \rho_\Lambda\left(\sum_{i=1}^d w_i X_i\right)\\
%\mathrm{s.t.} 
%\sum_{i=1}^d w_i = 1,\\ w^t\mu \geq r_\mathrm{min}, \\w_i\geq 0
%\end{align*}
%Reformulating in the standard form yields
\begin{align} \label{eq:optimisation}
\operatorname{min}_{\bfw\in\mathbb{R}^d}~~  -& \rho_\Lambda\left(\bfw^\top \bfX\right)\\ 
\text{such that} \quad\notag
& \one^\top \bfw = 1,\quad (  1 - \bfw^\top \bfmu / r_\mathrm{min} )\leq 0, \quad-\bfw\leq 0, 
\end{align}
where $\one^\top \bfw = \sum_{i=1}^d w_i$ and $\one = (1)_i$.

%We define $\R^d_0 = \{\bfw\in\R^d,\one^\top \bfw  =0\}$ and the affine space $\R^d_1 = \{\one\}\oplus \R^d_0 = \{\bfw\in\R^d,\one^\top \bfw  =1\}$. 

In order to solve \eqref{eq:optimisation}, we consider two practical reformulations, the first one based on a penalty method approach~\cite[Chapter 17]{nocedal:06} and the second one on the Karush-Kuhn-Tucker (KKT) conditions~\cite{kuhn:51}. We denote with $\R^d_1$ the affine subspace of elements $\bfw$ that sum to one, formally $\R^d_1 = \{\one\}\oplus \R^d_0 = \{\bfw\in\R^d,\one^\top \bfw  =1\}$, where $\mathbb{R}^d_0$ is the subspace of the elements $\bfw$ summing to zero, i.e. $\mathbb{R}^d_0 = \{\bfw \in \mathbb{R}^d \mid \mathbf{1}^\top \bfw = 0\}$. The two alternative formulations are given below.

\begin{method}[\textbf{Penalty approach}]

For $t > 0$, an approximated solution of \eqref{eq:optimisation} is given by
\begin{align*}
\bfw_{t, \mathrm{opt}} = \operatorname{arg\,amin}_{\bfw\in\mathbb{R}_1^d}~~  & - \rho_\Lambda\left(\bfw^\top \bfX\right)  + t/2\, (1 -  \bfw^\top\bfmu/r_\mathrm{min}  )_+^2 + t/2 \,\one^\top \left(-\bfw\right)_+^2,
\end{align*}
where $(x)_+ = \max\{x, 0\}$ (evaluated element-wise for a vector).  We set $t=100$, unless otherwise specified.   
\end{method} 
    
%This penalty approach is based on~\cite[Theorem 17.1]{nocedal:06}, which states that for a sequence of penalty parameters $(t_k)_{k\in\N}$, where $t_k\rightarrow \infty$ and such that the penalised minimisation problem has a finite solution for each $t_k$, every limit point of the sequence $(\bfw_{t_k, \mathrm{opt}})_t$ is a solution to~\eqref{eq:optimisation}. [HOW DO WE KNOW THAT IN OUR CASE THIS ALWAYS HAPPEN?] 

 \begin{method}[\textbf{KKT}]

On $\R^d_1 \times \R^d \times \R$, consider the Lagrangian
\[
\mathcal{L}(\bfw, \bflambdaw, \lambdar) = - \rho_\Lambda\left(\bfw^\top\bfX\right) - \bflambdaw^\top \bfw + \lambdar (1 - \bfw^\top\bfmu / r_\mathrm{min}).
\]
The solution of \eqref{eq:optimisation} is given by $(\bfw_{\mathrm{opt}}, \bflambdawopt, \lambdaropt)\in \R_1^d\times \R^d\times \R$, such that
\begin{align*}
\bfnabla_\bfw^0 \mathcal{L}(\bfw_\mathrm{opt}, \bflambdawopt, \lambdaropt) &= 0,\\
( 1 - \bfw_\mathrm{opt}^\top\bfmu /r_\mathrm{min}  )\leq 0 ,\quad&
-\bfw \leq 0\\
\lambdaropt\geq 0, \quad &
\bflambdawopt\geq 0\\
\lambdaropt  (1 - \bfw^\top\bfmu/r_\mathrm{min}) = 0, \quad &
\bfw^\top \bflambdawopt = 0.
\end{align*}
where $\bfnabla_\bfw^0 \mathcal{L}$ denotes the gradient of the Lagrangian on $\mathbb{R}^d_1$.
\end{method} 

% Optimisation method: Gradient descent mit Armijo, KKT vs penalty
We use a gradient descent by~\cite{curry:44} with a step-size based on Armijo's line-search \cite{armijo:66}, supported by the presented Newton method to optimise the lambda quantile over all possible portfolios. I.e. we update the weight as
$
w_{k+1} = w_k - \eta_0 2^{-j} \bfnabla_\bfw^0 g(\bfw_k),
$
for $g\colon \R^d_1\rightarrow \R$ being either the Lagrangian or the objective function of the penalty method and $\bfnabla_\bfw^0 g$ denoting the gradient with respect to the affine space. The initial step-size $\eta_0$ is set to $0.1$.
The actual step-size is determined by the Armijo rule, where $j$ is the smallest whole integer, such that
$
g( w_k - \eta_0 2^{-j}  \bfnabla_\bfw^0 g(\bfw_k)) - g(w_k)
\leq 
\eta_0 2^{-j} c_1 \| \bfnabla_\bfw^0 g(\bfw_k)  \|^2,
$ 
where we chose $c=0.1$.
This rule allows for controlling the step size and maintaining optimal gradient performance in case weights are rescaled. %[ADDED THIS SENTENCE]

In the KKT method we also update the Lagrange multipliers to

$
\lambda_\mathrm{w}^{k+1} = \left( \lambda_\mathrm{w}^k - 10^{-1}   w_k\right)_+$, and 
$\lambda_\mathrm{r}^{k+1} = \left( \lambda_\mathrm{r}^k - 10^{-1}  (r_\mathrm{min} - w_k^t\mu)/r_\mathrm{min} \right)_+
$.
In all cases, we stop the gradient descent once the norm of the gradient is smaller than a predefined tolerance (we use $0.001$ unless stated otherwise). In the case of the KKT method, also the inequalities and the complementarity conditions need to be fulfilled up to this accuracy.

Notice that, for a function $g\colon\R^d_1\rightarrow \R$ that extends differentiability onto $\R^d$, we can compute the gradient with respect to the affine space by projecting the gradient on $\R^d$ onto the tangent space $\R^d_0$. Hence, $\bfnabla_\bfw^0 g\colon \R^d_1\rightarrow \R^d_0$ is given by:
\[
\bfnabla_\bfw^0 g(\bfw) = \Pi_0 \bfnabla_\bfw g(\bfw ) 
= \bfnabla_\bfw g(\bfw)  - \frac{\one^\top\bfnabla_\bfw g(\bfw)}{\one^\top\one} \one.
\]

The use of a Newton method within a gradient descent results in nested iterations. We note that we can use the previous iteration of the lambda quantile as the starting value for the Newton iteration. Thanks to local quadratic convergence, this yields a small number of steps and an efficient evaluation of the lambda quantile in each iteration step. 

To evaluate the gradient of the lambda quantile $\bfnabla_\bfw \rho_\Lambda (\bfw)$, we use the result described in \cite{ince:22} and reported in formula \eqref{eq:LQgrad}. This requires additional assumptions on the distribution function of the portfolio random variable, as discussed in Remark \ref{rem:Ince}. A class of distributions smooth enough to satisfy the conditions of Remark \ref{rem:Ince} are the elliptical distributions, which include also the multivariate normal and the multivariate Student's t-distribution. These distributions are particularly relevant in portfolio management because they preserve key results of portfolio theory, such as separation theorem, rules of ordering under uncertainty, mutual fund separation theorems, and capital asset pricing model \cite{owen1983class}. However, alternative assumptions on the join distribution satisfying Remark \ref{rem:Ince} can be considered, including separately modelling marginals and dependences between returns, as long as a closed formula of the conditional expectation is available. In the following remark, we provide the formula for the gradient of the lambda quantile under this assumption.

\begin{remark}
Assume $\bfX=(X_i)_i$ is a d-dimensional random vector elliptically distributed $
\bfX \sim \mathcal{E}_d(\bfmu, \bfSigma)
$ with $\bfmu \in \mathbb{R}^d$, positive semidefinite matrix $\bfSigma\in\R^{d\times d}$ and density $\varphi$. The formula for the gradient of the lambda quantile can be derived as follows:
\[
\bfnabla_\bfw \rho_\Lambda (\bfw) = 
\frac{\varphi_{\bfw}(\rho_\bfw)}{\varphi_{\bfw}(\rho_\bfw) - \Lambda'(\rho_\bfw)} \left( \bfmu + \frac{\rho_\bfw- \mu_\bfw}{\sigma_\bfw^2}\bfw^\top\bfSigma\right),
\]
where $\rho_\bfw$ and $\varphi_{\bfw}$ are respectively the lambda quantile and the probability density function of the portfolio $\bfw^\top\bfX$. For this result, we use the conditional distribution derived in~\cite[Corollary 5]{cambanis1981theory}.
\end{remark}

\subsection{Practical examples}

In this chapter, we examine practical examples of portfolio optimisation using lambda quantiles while assuming multivariate normal and multivariate t distributions of the assets. In particular, if $\bfX$ follows a multivariate normal distribution, $\bfX\sim\mathcal{N}(\bfmu, \bfSigma)$, the portfolio is normally distributed $\bfw^\top\bfX\sim \mathcal{N}(\mu_\bfw, \sigma_\bfw^2)$. In case of a multivariate t distribution of the assets $\bfX\sim t_\nu(\bfmu, \bfSigma)$, for $\nu > 0$, the portfolio follows a non-standardized t distribution $\bfw^\top\bfX \sim t_\nu(\mu_\bfw, \sigma_\bfw^2)$.

In all three examples, we consider a piecewise linear lambda function, for $0<\lambda_m<\lambda_M<1$ and $x_m < x_M$:
\[
\Lambda(x) = \begin{cases}
\lambda_m, \quad & x<x_m,\\
\lambda_m + \frac{x  - x_m}{x_M - x_m} (\lambda_M - \lambda_m), \quad & x_m \leq x < x_M,\\
\lambda_M, \quad & x\geq x_M.\\
\end{cases}
\]
The values $x_m, x_M, \lambda_m, \lambda_M$ can be chosen reasonably, ensuring unique solvability of~\eqref{eq:lq}. In the following, we take $\lambda_m = 0.025$, $\lambda_M = 0.05$, $x_m = -0.257$ and $x_M = 0.277$. Furthermore, we assume that the required minimal return is $r_\mathrm{min} = 0.015$.

\subsubsection{Two-dimensional optimisation}
In our first example, we consider the returns $X_1\sim\mathcal{N}(0.01, 0.1^2)$, $X_2\sim\mathcal{N}(0.02, 0.15^2)$ with $\operatorname{corr}(X_1, X_2) = 0.4$.  
We use the initial weights $w_\mathrm{init} = (0.1, 0.9)^\top$.

\begin{figure}
    \centering
    \includegraphics[width=.5\textwidth]{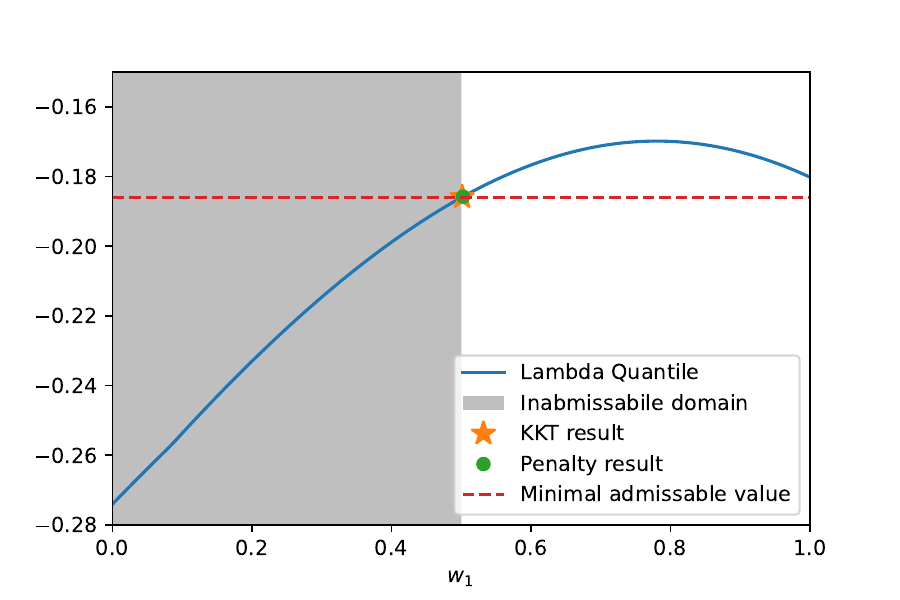}
    \caption{Plot of the two asset optimisation problem. Showing the two numerical solutions and the minimal lambda quantile. }
    \label{fig:two_assets_plot}
\end{figure}

Both methods converge within the required tolerance and provide similar solutions. The results are visualised in Figure~\ref{fig:two_assets_plot}.
The optimal values are compared in Table~\ref{tab:results_two_assets}.

Additionally, we examine the convergence of the algorithms by comparing the number of gradient descent steps, the required calls to the\ouralgorithm{} method, and the number of\ouralgorithm{} steps, as shown in Table~\ref{tab:convergence_two_assets}. We observe that while the penalty method requires fewer steps to find a solution, it requires more calls to the\ouralgorithm{} method. The performance of the\ouralgorithm{} is efficient and comparable for both methods, with each requiring approximately two\ouralgorithm{} steps per call to converge. 

\begin{table}\small
    \centering
        \begin{tabular}{@{}lllll@{}}\toprule
           & $w_1$ & $w_2$& $\bfw^\top \mu$& $\rho_\bfw$  \\ \midrule
        Exact solution & $0.5$ & 0.5 & 1.5\% & $-0.186040$\\
        Penalty & 0.5025 & 0.4975 & 1.4975\% & $-0.185762$\\
        KKT & 0.5014  & 0.4986 & 1.4986\%& $-0.185882$\\\bottomrule
    \end{tabular}
    \caption{Optimisation results for two assets}
    \label{tab:results_two_assets}
\end{table}

\begin{table}\small
    \centering
    \begin{tabular}{@{}lccccccc@{}}\toprule
           &  \# steps g.d.   &  & \multicolumn{2}{c}{\# calls  \ouralgorithm{}} &  & \multicolumn{2}{c}{\# steps \ouralgorithm{}}  \\
         \cmidrule(lr){4-5} \cmidrule(lr){7-8}
         & & & total & relative & & total & relative  \\ \midrule
         Penalty & 7 && 31 & 3.88 && 67 & 2.16\\
         KKT & 20 && 21 & 1.00 && 52 & 2.48
         \\\bottomrule
    \end{tabular}
    \caption{Overview of the iterative optimisation in the two-asset case. Number of gradient decent steps, number of calls to the\ouralgorithm{} algorithm (total number and relative to the number of gradient steps plus the initial evaluation) and total number of iterations in the\ouralgorithm{} algorithm (total and relative to the number of calls).  }
    \label{tab:convergence_two_assets}
\end{table}

\subsubsection{Three asset optimisations with normal distribution}
Consider $\bfX\sim\mathcal{N}(\bfmu, \bfSigma)$ where $\bfmu = (0.013, 0.014, 0.02)^\top$, $\bfSigma = (\bfsigma^\top \bfsigma) \odot \bfrho $ with $\bfsigma = (0.13, 0.0145, 0.15)^\top$ and $\bfrho = \begin{pmatrix}
1 & 0.4& 0.08\\
0.4& 1& 0.2\\
0.08 & 0.2 & 1
\end{pmatrix}$. We take $w_\mathrm{init} = (0.1, 0.5, 0.4)^\top$ as initial weights.

As a robustness check, in this example, we investigate the convergence for varying levels of accuracy. Specifically, we set the tolerance of the gradient descent method %[IS NOT THE TOLERANCE OF THE GRADIENT DESC METHOD? - Indeed that is correct] 
to 1e$-2$, 1e$-3$ and 1e$-4$. For the penalty method, we also adapt the penalty parameter $t$ to 10, 100 and 1000, respectively. We note that the tolerance of the\ouralgorithm{} method remains unchanged at 1e$-8$ in all cases. 
The results are shown in Table~\ref{tab:results_three_assets_new_idea}. We observe reasonable results, except for the penalty method with the lowest accuracy (tol=0.01), where the return is $1.4381\%$, significantly below the minimal return of $1.5\%$. With smaller tolerances, the constraints are observed more accurately, and both algorithms converge to the same solution.

In Table~\ref{tab:convergence_three_assets_new_idea} we report the number of steps needed for the gradient descent and the\ouralgorithm{} method. As expected, we observe a higher number of gradient descent steps for smaller tolerances, which results in a growing number of calls to the\ouralgorithm{} algorithms. Consistently, the penalty method requires fewer gradient steps than the KKT method, which also needs to satisfy the Lagrangian multiplier conditions. However, as in the previous example, the penalty method requires a greater number of calls to the\ouralgorithm{} algorithm. This is likely due to the large penalty factor, which influences the condition of the underlying optimisation problem. In all cases, we observe a stable number of\ouralgorithm{} steps per function calls. On average, 2-3 steps are required to compute the lambda quantile to the required accuracy of $10^{-8}$, confirming the algorithm's robustness.

\begin{table}\small
    \centering
        \begin{tabular}{@{}rlllll@{}}\toprule
           & $w_1$ & $w_2$&  $w_3$& $\bfw^\top \mu$& $\rho_\bfw$  \\ \midrule
           tol $=0.01$ \\
           Penalty & $-0.007663$ & $0.945496$ & $0.062168$ & $1.4381$\% & $-0.017484$ \\
           KKT & $0.002748$ & $0.823989$ & $0.173264$ & $1.5037$\% & $-0.040684$ \\
           tol $=0.001$ \\
        Penalty & $-0.000793$ & $0.848402$ & $0.152391$ & $1.4915$\% & $-0.035719$ \\
        KKT & $0.000495$ & $0.830615$ & $0.168891$ & $1.5013$\% & $-0.039569$ \\
           tol $=0.0001$ \\
        Penalty & $-0.000079$ & $0.834874$ & $0.165205$ & $1.4991$\% & $-0.038689$ \\
           KKT & $-0.000029$ & $0.833449$ & $0.166580$ & $1.5000$\% & $-0.039010$ \\
        \bottomrule
    \end{tabular}
    \caption{Optimisation results for three assets with a multivariate normal distribution}
    \label{tab:results_three_assets_new_idea}
\end{table}

\begin{table}\small
    \centering
    \begin{tabular}{@{}lccccccc@{}}\toprule
           &  \# steps g.d.   &  & \multicolumn{2}{c}{\# calls  \ouralgorithm{}} &  & \multicolumn{2}{c}{\# steps \ouralgorithm{}}  \\
         \cmidrule(lr){4-5} \cmidrule(lr){7-8}
         & & & total & relative & & total & relative  \\ \midrule
           tol $=0.01$ \\
           Penalty & 8 && 23 & 2.56 && 56 & 2.43\\
           KKT & 21 && 22 & 1.00 && 62 & 2.82\\
           tol $=0.001$ \\
         Penalty & 19 && 101 & 5.05 && 234 & 2.32\\
         KKT & 45 && 46 & 1.00 && 117 & 2.54\\
           tol $=0.0001$ \\
           Penalty & 45 && 381 & 8.28 && 697 & 1.83\\
           KKT & 78 && 79 & 1.00 && 179 & 2.27\\
         \\\bottomrule
    \end{tabular}
    \caption{Overview of the iterative optimisation in the three-asset case. Number of gradient decent steps, number of calls to the\ouralgorithm{} algorithm (total number and relative to the number of gradient steps plus the initial evaluation) and total number of iterations in the\ouralgorithm{} algorithm (total and relative to the number of calls).  }
    \label{tab:convergence_three_assets_new_idea}
\end{table}

\subsubsection{Three asset optimisation with multivariate t distribution}
In this section, we consider three different situations with various levels of active constraints.

\begin{table}\small
    \centering
        \begin{tabular}{@{}rlllll@{}}\toprule
           & $w_1$ & $w_2$&  $w_3$& $\bfw^\top \mu$& $\rho_\bfw$  \\ \midrule
           unconstrained solution\\
           Penalty & 0.421394 & 0.224291 & 0.354315 & 1.5704\% & $-0.293923$ \\
           KKT & 0.421968 & 0.223964 & 0.354067 & 1.5702\% & $-0.293923$ \\
        one active constraint \\
        Penalty & 0.614475 & 0.115759 & 0.269766 & 1.5004\% & $-0.254279$ \\
        KKT & 0.613228 & 0.116176 & 0.270597 & 1.5010\% & $-0.254279$ \\
        two active constaints\\
           Penalty & -0.001257 & 0.857182 & 0.144075 & 1.4866\% & $-0.060354$ \\
           KKT & 0.000733 & 0.831995 & 0.167273 & 1.5003\% & $-0.069013$ \\
        \bottomrule
    \end{tabular}
    \caption{Optimisation results for three assets with a multivariate t distribution}
    \label{tab:results_t_distribution_new_idea}
\end{table}

Assume $\bfX\sim t_\nu(\bfmu, \bfSigma)$ where $\nu=3$, $\bfmu = (0.013, 0.014, 0.02)^\top$, $\bfSigma = (\bfsigma^\top \bfsigma) \odot \bfrho $ with $\bfrho = 
\begin{pmatrix}
1 & 0.4& 0.08\\
0.4& 1& 0.2\\
0.08 & 0.2 & 1
\end{pmatrix}$. To explore the performance of the algorithm under different levels of activation of the constraints, we set $\bfsigma$ as follows:
\begin{itemize}
    \item Unconstrained: $\bfsigma = (0.13, 0.0145, 0.15)^\top$ 
    \item One active constraint:  $\bfsigma = (0.1, 0.145, 0.15)^\top$
    \item Two active constraints: $\bfsigma = (0.13, 0.0145, 0.15)^\top$
\end{itemize}

We consider $w_\mathrm{init} = (0.1, 0.5, 0.4)^\top$ as initial weights.

Results are in Table~\ref{tab:results_t_distribution_new_idea} and~\ref{tab:convergence_t_distribution_new_idea}. As expected, the more constraints are active, the more gradient descent steps are required to find a solution. Similarly to previous examples, the penalty method requires fewer steps than KKT method. However, the\ouralgorithm{} consistently performs with an average of 2-3 steps per function call.

\begin{table}\small
    \centering
    \begin{tabular}{@{}rccccccc@{}}\toprule
           &  \# steps g.d.   &  & \multicolumn{2}{c}{\# calls  \ouralgorithm{}} &  & \multicolumn{2}{c}{\# steps \ouralgorithm{}}  \\
         \cmidrule(lr){4-5} \cmidrule(lr){7-8}
         & & & total & relative & & total & relative  \\ \midrule
         unconstrained solution \\
         Penalty & 4 && 5 & 1.00 && 16 & 3.20\\
         KKT & 12 && 13 & 1.00 && 42 & 3.23\\
         one active constraint \\
Penalty & 5 && 6 & 1.00 && 18 & 3.00\\
KKT & 12 && 13 & 1.00 && 34 & 2.62\\
two active constraint \\ 
Penalty & 25 && 150 & 5.77 && 375 & 2.50\\
KKT & 32 && 42 & 1.27 && 117 & 2.79\\
         \bottomrule
    \end{tabular}
    \caption{Overview of the iterative optimisation in the three-asset case with a multivariate t distribution. Number of gradient decent steps, number of calls to the\ouralgorithm{} algorithm (total number and relative to the number of gradient steps plus the initial evaluation) and total number of iterations in the \ouralgorithm{} algorithm (total and relative to the number of calls).  }
    \label{tab:convergence_t_distribution_new_idea}
\end{table}

\section{Conclusion}\label{sec:conclusion}

We investigated the development of a robust and accurate numerical method for computing lambda quantiles, with particular emphasis on applications in portfolio selection. Our findings demonstrate that lambda quantiles offer significantly greater flexibility than classical risk measures, such as Value at Risk, while maintaining solid computational performance.

We proposed an algorithm for the rapid computation of lambda quantiles that guarantees convergence under very mild assumptions, including cases involving discontinuous functions, while preserving elicitability. The algorithm is inspired by the Brent method but differs by combining the reliability of the bisection method with the fast convergence properties of Newton iterations, rather than relying on the secant method used in Brent’s approach. The bisection method ensures convergence, while Newton’s method contributes to fast local convergence, exhibiting a quadratic rate under local differentiability assumptions for single roots. In this setting, Newton’s algorithm offers the additional advantage of solving the equation using the derivative of the distribution function - i.e., the density -which is typically easier to compute than the distribution function itself. 

Numerical experiments support the theoretical results, with convergence observed in all tested cases. Local quadratic convergence occurred consistently, except when the lambda quantile coincided with a discontinuity in either the distribution or the lambda function, as predicted by the theory. In such cases, we recommend an interval analysis approach.

Finally, we applied the method to a standard portfolio optimisation problem using lambda quantiles. Constraints were implemented using both a KKT-based formulation and a penalty method. Building on prior work establishing sensitivity formulas for lambda quantiles, we evaluated the solver's performance at each optimisation step. The algorithm required only a few inner iterations per step, resulting in rapid overall convergence. These results confirm not only the computational tractability of lambda quantiles but also their practical utility in optimisation tasks.

\bibliographystyle{plain}
\bibliography{lit.bib}

@book{nesterov:18,
title = {Lectures on Convex Optimization},
author = {Yurii Nesterov},
series = {Springer Optimization and Its Applications},
edition = {2nd},
publisher = {Springer},
year = {2018}
}

@book{quarteroni:07,
title = {Numerical Mathematics},
author = {Alfio Quarteroni and Riccardo Sacco and Fausto Saleri},
series = {Texts in Applied Mathematics},
publisher = {Springer},
year = {2007}
}

@article{sorensen:82trustregionnewton,
author = {Sorensen, D. C.},
title = {Newton’s Method with a Model Trust Region Modification},
journal = {SIAM Journal on Numerical Analysis},
volume = {19},
number = {2},
pages = {409-426},
year = {1982}
}

@book{moore:09,
title = {Introduction to Interval Analysis},
author= {Ramon E. Moore and R. Baker Kearfott and Michael J. Cloud},
year = {2009},
publisher = {SIAM}
}

@article{brent:71,
  title={An algorithm with guaranteed convergence for finding a zero of a function},
  author={Brent, Richard P.},
  journal={The Computer Journal},
  volume={14},
  number={4},
  pages={422--425},
  year={1971},
  publisher={Oxford University Press}
}

@article{balakrishnan:85,
 author = {N. Balakrishnan and S. Kocherlakota},
 journal = {Sankhyā: The Indian Journal of Statistics, Series B},
 number = {2},
 pages = {161--178},
 title = {On the Double {Weibull} Distribution: Order Statistics and Estimation},
 volume = {47},
 year = {1985}
}

@book{corneujols:18,
title = {Optimization Methods in Finance},
author = {Gérard Cornuéjols and Javier Peña and Reha Tütüncü},
year = {2018},
publisher = {Cambridge University Press}
}

@book{nocedal:06,
title = {Numerical Optimization},
author = {Jorge Nocedal and Stephen J. Wright},
year = {2006},
edition = {2nd},
publisher = {Springer},
location = {New York},
}

@article{ince:22,
author = {Akif Ince and Ilaria Peri and Silvana Pesenti},
title = {Risk contributions of lambda quantiles},
journal = {Quantitative Finance},
volume = {22},
number = {10},
pages = {1871-1891},
year = {2022},
publisher = {Routledge},

}

@book{deuflhard:11,
author = {Peter Deuflhard},
title = {Newton Methods for Nonlinear Problems},
subtitle = {Affine Invariance and Adaptive Algorithms},
publisher = {Springer},
location = {Berlin},
year = {2011}
}

@article{armijo:66,
author = {Larry Armijo},
title = {{Minimization of functions having {Lipschitz} continuous first partial derivatives}},
volume = {16},
journal = {Pacific Journal of Mathematics},
number = {1},
publisher = {Pacific Journal of Mathematics, A Non-profit Corporation},
pages = {1-3},
year = {1966},
}

@article{curry:44,
  title={The method of steepest descent for non-linear minimization problems},
  author={Haskell B. Curry},
  journal={Quarterly of Applied Mathematics},
  year={1944},
  volume={2},
  pages={258-261},
}

@article{kuhn:51,
year = {1951},
title = {Nonlinear Programming},
author = {H. W. Kuhn, A. W. Tucker},
journal = {Berkeley Symp. on Math. Statist. and Prob.},
pages = {481-492}
}

@article{bp19,
	author = {Bellini, Fabio and Peri, Ilaria},
	journal = {SIAM Journal on Financial Mathematics},
    volume = {13},
    number = {1},
	title = {An axiomatization of {$\Lambda$}-quantiles},
	year = {2022}}

@article{bpr17,
	author = {Burzoni, Matteo and Peri, Ilaria and Ruffo, Chiara M},
	journal = {Quantitative Finance},
	number = {11},
	pages = {1735--1743},
	publisher = {Taylor \& Francis},
	title = {On the properties of the Lambda value at risk: robustness, elicitability and consistency},
	volume = {17},
	year = {2017}}

@article{hmp18,
	author = {Hitaj, Asmerilda and Mateus, Cesario and Peri, Ilaria},
	journal = {Risks},
	number = {1},
	pages = {17},
	publisher = {Multidisciplinary Digital Publishing Institute},
	title = {Lambda value at risk and regulatory capital: a dynamic approach to tail risk},
	volume = {6},
	year = {2018}}

@article{fmp14,
	author = {Frittelli, Marco and Maggis, Marco and Peri, Ilaria},
	fjournal = {Mathematical Finance. An International Journal of Mathematics, Statistics and Financial Economics},
	issn = {0960-1627},
	journal = {Mathematical Finance},
	mrclass = {91B30},
	mrnumber = {3274936},
	mrreviewer = {Shuanming Li},
	number = {3},
	pages = {442--463},
	title = {Risk measures on {$\mathcal{P}(\mathbb{R})$} and value at risk with probability/loss function},
	volume = {24},
	year = {2014}}

@article{liu2024risk,
  title={Risk sharing with Lambda value at risk},
  author={Liu, Peng},
  journal={Mathematics of Operations Research},
  year={2024},
volume = {50},
number = {1}
}

@article{han2021cash,
  title={Cash-subadditive risk measures without quasi-convexity},
  author={Han, Xia and Wang, Qiuqi and Wang, Ruodu and Xia, Jianming},
  journal={Mathematics of Operations Research},
  year={2025},
volume={published online}
}

@article{corbetta2018backtesting,
  title={Backtesting lambda value at risk},
  author={Corbetta, Jacopo and Peri, Ilaria},
  journal={The European Journal of Finance},
  volume={24},
  number={13},
  pages={1075--1087},
  year={2018},
  publisher={Taylor \& Francis}
}

@article{gneiting2011making,
  title={Making and evaluating point forecasts},
  author={Gneiting, Tilmann},
  journal={Journal of the American Statistical Association},
  volume={106},
  number={494},
  pages={746--762},
  year={2011},
  publisher={Taylor \& Francis}
}

@article{bignozzi2020risk,
  title={Risk measures based on benchmark loss distributions},
  author={Bignozzi, Valeria and Burzoni, Matteo and Munari, Cosimo},
  journal={Journal of Risk and Insurance},
  volume={87},
  number={2},
  pages={437--475},
  year={2020},
  publisher={Wiley Online Library}
}

@article{balbas2023lambda,
  title={Lambda-quantiles as fixed points},
  author={Balb{\'a}s, Alejandro and Balb{\'a}s, Beatriz and Balb{\'a}s, Raquel},
  journal={Available at SSRN 4583950},
  year={2023}
}

@article{salo2023fifty,
  title={Fifty years of portfolio optimization - {A European} perspective},
  author={Salo, Ahti and Doumpos, Michalis and Liesi{\"o}, Juuso and Zopounidis, Constantin},
  journal={European Journal of Operational Research},
  year={2023},
  publisher={Elsevier}
}

@article{kolm201460,
  title={60 years of portfolio optimization: Practical challenges and current trends},
  author={Kolm, Petter N and T{\"u}t{\"u}nc{\"u}, Reha and Fabozzi, Frank J},
  journal={European Journal of Operational Research},
  volume={234},
  number={2},
  pages={356--371},
  year={2014},
  publisher={Elsevier}
}

@article{bellini2015elicitable,
  title={On elicitable risk measures},
  author={Bellini, Fabio and Bignozzi, Valeria},
  journal={Quantitative Finance},
  volume={15},
  number={5},
  pages={725--733},
  year={2015},
  publisher={Taylor \& Francis}
}

@article{owen1983class,
  title={On the class of elliptical distributions and their applications to the theory of portfolio choice},
  author={Owen, Joel and Rabinovitch, Ramon},
  journal={The Journal of Finance},
  volume={38},
  number={3},
  pages={745--752},
  year={1983},
  publisher={Wiley Online Library}
}

@article{cambanis1981theory,
  title={On the theory of elliptically contoured distributions},
  author={Cambanis, Stamatis and Huang, Steel and Simons, Gordon},
  journal={Journal of Multivariate Analysis},
  volume={11},
  number={3},
  pages={368--385},
  year={1981},
  publisher={Elsevier}
}

@article{xia2024optimal,
  title={Optimal risk sharing for lambda value-at-risk},
  author={Xia, Zichao and Hu, Taizhong},
  journal={Advances in Applied Probability},
  pages={1--33},
  year={2024},
  publisher={Cambridge University Press}
}

@article{liu2024brisk,
  title={Risk sharing with Lambda value at risk under heterogeneous beliefs},
  author={Liu, Peng and Tsanakas, Andreas and Wei, Yunran},
  journal={arXiv preprint arXiv:2408.03147},
  year={2024}
}

@article{han2024robust,
  title={Robust {$\Lambda$}-quantiles and extremal distributions},
  author={Han, Xia and Liu, Peng},
  journal={Mathematical Finance},
  year={2025}
}

@article{boonen2024optimal,
  title={Optimal insurance design with Lambda-Value-at-Risk},
  author={Boonen, Tim J and Chen, Yuyu and Han, Xia and Wang, Qiuqi},
  journal={European Journal of Operational Research},
volume = {327},
number = {1},
pages ={232-246},
  year={2025}
}

@article{chambers2024max,
  title={Max- and min-stability under first-order stochastic dominance},
  author={Chambers, Christopher and Miller, Alan and Wang, Ruodu and Wu, Qinyu},
  journal={Mathematics and Financial Economics },
volume = {19},
pages = {641-659},
  year={2025}
}

\begin{appendix}
\section{Deferred Proofs} \label{sec:appendix_proofs}
\subsection{Proof of Theorem~\ref{thm:convergence}}
\label{sec:proof_of_theorem_1}
As for $\bar x = x_l$ or $\bar x = x_r$ the algorithm immediately terminates successfully, we can consider $\bar x \in (\xmin, \xmax)$ in the following.

We formally consider the infinite sequence of iterates created by Algorithm~\ref{alg:newton} for $\varepsilon=0$ and $N_{\max{}} = \infty$ . 
If this sequence converges, it will eventually fulfil one of the two exit criteria, $|F(x_0) - L(x_0)| \leq \varepsilon$ or $|x_0 - \bar x| \leq |x_r - x_l| < \varepsilon$, and the algorithm successfully finished. For $N_{\max{}}$ large enough, the algorithm does not finish earlier either. 

We denote the state of all variables after the $i$-th iteration of the algorithm with a superscript $i$, in particular $x_l^i, x_r^i$ and $x_i = x_0^i$.
We note that the algorithm either continues using a standard Newton step (line~\ref{line:newton}) or a bisection step (line~\ref{line:bisection}).

By construction, $\bar x \in [x_l^i, x_r^i] \subset [\xmin, \xmax]$ holds for all $i>0$. Thus if the length of the interval converges to zero, $x_r^i - x_l^i \rightarrow 0$, we have convergence $x_i\rightarrow \bar x$. 

If the length of the interval does not converge to zero, there must be an index $N$, such that no bisection steps have been taken for $i>N$. Consequently for $i>N$, the iterates $x_i$ update following the standard Newton method.
By construction the sequence $(x_i)_i$ is bounded, so
\[
\xinf = \lim\inf x_i, \text{ and } \xsup = \lim\sup x_i
\]
are both finite.

If $\xinf=\xsup$, the sequence $x_i$ converges and 
\[
\frac{|f(x_i)|}{|f'(x_i)|}\rightarrow 0.
\]
As $f'$ is càdlàg by Assumption~\ref{assumption1}, it is bounded on  $[\xmin, \xmax]$, and hence $f(x_i) \rightarrow 0$. 
If a subsequence of $x_i$ approaches $\xinf$ from the right, the right-continuity of $f$ yields $f(\xinf) = 0$ and hence $\xinf=\bar x$. In the other case, a subsequence of $x_i$ approaches $\xinf$ from the left and hence $f(\xinf-) = 0$. By Assumption~\ref{assumption1} also in this case $\xinf = \bar x$.

Now assume $\xinf \neq \xsup$, i.e. $\tau = \xsup -\xinf > 0$. By construction $F(\xinf) - L(\xinf) \leq 0$ and $F(\xsup) - L(\xsup) \geq0$ and for all $i\in\N$,
$\xinf, \xsup \in [x_l^i, x_r^i]$.
As either $x_i = x_l^{i+1}$ or $x_i = x_r^{i+1}$ by construction, this implies that $x_i\not\in(\xinf, \xsup)$.

By definition, there exist $i_l$, such that $|x_{i_l}-\xinf|\leq \delta \tau$ and $i_r$, such that $|x_{i_r} - \xsup|\leq \delta \tau$. For all further iterates $i>\max\{i_l, i_r\}$ it holds $x_i\in[\xinf-\delta \tau, \xinf] \cup [\xsup, \xsup+\delta \tau ]$ (with infinite iterates in each of the sub-intervals), $x_l^i \in [\xinf-\delta \tau, \xinf] $, $x_{r}^i \in[\xsup, \xsup+\delta \tau ]$.% and  $x_r^i-x_l^i\geq \tau$.

As $\xinf\neq\xsup$, there exists $\hat \imath >  \max\{i_l, i_r\}$, such that $x_{\hat i} \in [\xinf-\delta \tau , \xinf] $ and $x_{\hat \imath+1} \in  [\xsup, \xsup+\delta \tau ]$. This implies $dx^{\hat \imath+1} = x_{\hat \imath + 1} - x_{\hat \imath} \geq \xsup - \xinf = \tau$ and $x_{\hat\imath+1} \geq \xsup \geq x_r^{\hat\imath+1} - \delta \tau  \geq x_r^{\hat\imath+1} - \delta |dx^{\hat \imath+1}| $. This means that the condition in line~\ref{line:bisection_condition} of Algorithm~\ref{alg:newton} is true and a bisection step is taken, which contradicts our assumptions. 
This concludes global convergence of the algorithm to the unique lambda quantile $\bar x$. 
\subsection{Proof of Theorem~\ref{thm:quadratic_convergence}}

As Theorem~\ref{thm:convergence} shows that the iteration converges, the iterates eventually come close enough to the solution that the Newton method converges quadratically. It remains to show that only a finite number of bisection steps are taken. If $x_i=\bar x$, the statement trivially holds, so we assume $|\bar x - x_i|>0$ for all $i>0$.

We first determine the region of local convergence using a Taylor series. 
For any $ \tau>0$ we define $I_\tau = [\bar x - \tau, \bar x + \tau]$ and 
\[
 C(\tau) = \frac{\sup_{x\in I_\tau}|f''(x)|}{2\inf_{x\in I_\tau}|f'(x)|}.
\]
As $f$ is locally $C^2$ and $f'(\bar x) \neq 0$, $C(\tau)$ is finite for sufficiently small $\tau$. 
We note furthermore that $C(\tau)$ is monotonic increasing in $\tau$. This implies that we can choose a $\tau$ sufficiently small, such that 
\begin{equation}\label{eq:condition_c}
(1+C(\tau)\tau)/2 < 1-\delta.
\end{equation}
We also choose $\tau$ sufficiently small such that we have $\xmin, \xmax\not\in I_\tau$, noting that the algorithm immediately returns with the solution if $\xmin= \bar x$ or $\xmax=\bar x$.
We note that this implies $C(\tau)\tau < 1$ and we denote $C=C(\tau)$ from now.

With $f\in C^2(I_\tau)$  we can apply Taylor's theorem on $f$ around $x$ for any $x\in I_\tau$ to show that
$0 = f(\bar x) = f(x) +  f'( x)  (\bar x - x) + \frac{f''(\xi)}{2} (x-\bar x)^2$ for some $\xi\in I_\tau$.
This shows quadratic convergence of a Newton algorithm on the interval $I_\tau$:
\begin{equation}\label{eq:proof_newton_quadratic_convergence}
 \left| \frac{f(x)}{f'(x)} + \bar x  - x \right| = \left|\frac{f''(\xi)}{2f'(x)} (x-\bar x)^2 \right| \leq C \left| x-\bar x\right|^2 \leq |x-\bar x| .
\end{equation}

Using the convergence shown in Theorem~\ref{thm:convergence}, there exists $n\in\N$, such that for $i>n$: $x_i \in I_\tau$ and~\eqref{eq:proof_newton_quadratic_convergence}  holds. Here, we use the same notation as in~\ref{sec:proof_of_theorem_1}, where $x_i$ denotes the iterate $x_0$ as defined at the end of the $i$-th iteration of the loop. 

In line~\ref{line:newton} of the algorithm, the Newton update is computed as $dx^{i+1} = -\frac{f(x_i)}{f'(x_i)}$ and 
it remains to show that 
\begin{equation}
x_i + dx^{i+1} \in (x_l^{i+1} + \delta |dx^{i+1}|, x_r^{i+1} - \delta|dx^{i+1}|), \label{cond:bisection}
\end{equation}
for all $i>n$ which implies that no bisection step is taken.
Assume this is not the case and consider the first instance of an iterate $x_i\in I_\tau$,  where~\eqref{cond:bisection} is not fulfilled.  Let us assume $f(x_i) < 0$, hence $x_l^{i+1} = x_i$, and note that an analogue calculation concludes the case $f(x_i) > 0$. 
Note that quadratic convergence in $I_\tau$~\eqref{eq:proof_newton_quadratic_convergence} implies that $|\bar x - x_i| \leq |\bar x - x_j|$ for $j<i$. As we also have $|\bar x - x_i| \leq \tau < |\bar x - \xmax|$ we have $|\bar x - x_i| \leq |\bar x - x_r^{i+1}|$. This is due to the fact that if $x_r^{i+1}$ is not a previous iterate $x_j$, $j < i$ it is still the initial value $\xmax$ and by construction $ \xmax\not\in I_\tau$. As a result $|\bar x - x_i|\leq  |x_r^{i+1}-x_l^{i+1}|/2$.

We recall from~\eqref{eq:proof_newton_quadratic_convergence} that 
$|x_i + dx^{i+1} - \bar x| \leq C (x_i - \bar x)^2 \leq C \tau |x_i - \bar x|$,
hence with ~\eqref{eq:condition_c}
\begin{align*}
|dx^{i+1}| &\leq (1 + C \tau) |x_i - \bar x| \leq (1+C\tau)/2  |x_r^{i+1}-x_l^{i+1}|\\
&      < (1-\delta) |x_r^{i+1}-x_l^{i+1}| <  |x_r^{i+1}-x_l^{i+1}|.
\end{align*}

Recalling $x_i = x_l^{i+1}$, we note that $dx^{i+1}>0$ and $x_i + dx^{i+1} > x_l^{i+1} + \delta dx^{i+1}$ since $\delta < 1$. On the other hand   using
$dx^{i+1}<  (1-\delta) |x_r^{i+1}-x_l^{i+1}|$ and $x_i = x_r^{i+1} -  |x_r^{i+1}-x_l^{i+1}|$, we have
\[x_i + dx^{i+1} < x_r^{i+1} - \delta |x_r^{i+1}-x_l^{i+1}| < x_r^{i+1} - \delta dx^{i+1},\] contradicting our assumption that~\eqref{cond:bisection} was not fulfilled.

The analogue argument for the case $x_i = x_r^{i+1}$ concludes the proof. Here $dx^{i+1}<0$ and $x_i + dx^{i+1} < x_r^i - \delta |dx^{i+1}|$ since $\delta < 1$. Similarly we have
\[x_i + dx^{i+1} > x_l^i + \delta |x_r^i-x_l^i| > x_l^i + \delta |dx^{i+1}|,\]
which concludes that $x_{i+1} = x_i + dx^{i+1}$ and for $i>n$ the algorithm only performs Newton steps.

The rest of the proof is standard, defining $e_i = |x_i - \bar x|$, we have
\[
e_{i+1} = \left|\bar x  - \left(x_i -  \frac{f(x_i)}{f'(x_i)}\right)\right| \leq C (x_i-\bar x)^2 = C e_i^2,
\]
with $C e_{i} < C\tau<1$, which shows quadratic convergence of the method for $i>N$. 

\end{appendix}

\end{document}